\documentclass{IEEEtran}

\usepackage{color}
\usepackage{cuted}
\usepackage{amsfonts}
\usepackage{mathrsfs}
\usepackage{amssymb,amsmath}
\usepackage[noblocks]{authblk}
\usepackage[compress]{cite}
\usepackage{bm}
\usepackage{algorithm}
\usepackage{algorithmic}
\usepackage{amsmath,amssymb,epsfig,graphics,subfigure}
\usepackage{theorem}
\usepackage[T1]{fontenc}
\usepackage[compress]{cite}
\usepackage{algorithm}
\usepackage{algorithmic}
\usepackage{stfloats}
\usepackage{subeqnarray}
\usepackage{cases}
\usepackage{subfigure}
\allowdisplaybreaks

\begin{document}

\title{A Framework on Complex Matrix Derivatives with Special Structure Constraints for Wireless Systems }
\author{Xin Ju, Shiqi Gong, Nan Zhao, \textsl{Senior Member}, \textsl{IEEE}, Chengwen Xing, 
Arumugam Nallanathan, \textsl{Fellow}, \textsl{IEEE}, and Dusit Niyato, \textsl{Fellow}, \textsl{IEEE}
\thanks{
X. Ju and C. Xing are with the School of Information and Electronics, 
Beijing Institute of Technology, Beijing 100081, China 
(E-mails: xinjubit@gmail.com; chengwenxing@ieee.org).

S. Gong is with the School of Cyberspace Science and Technology, Beijing Institute of Technology, Beijing 100081, China
(E-mail: gsqyx@163.com).

N. Zhao is with  the School
of Information and Communication Engineering, Dalian University of
Technology, Dalian 116024, China (E-mail: zhaonan@dlut.edu.cn).

Arumugam Nallanathan is with the QueenMary University of London, E14NS
London, U.K. (E-mail: a.nallanathan@qmul.ac.uk).

Dusit Niyato is with the School of Computer Science and Engineering,
Nanyang Technological University, Singapore 639798 (E-mail: dniyato@
ntu.edu.sg).

\vspace*{-4mm}
} %
\vspace*{-5mm}
}
\maketitle

\vspace{-20mm}

\begin{abstract}
  Matrix-variate optimization plays a central role in advanced wireless system designs. In this paper, we aim to explore optimal solutions of matrix variables under two special structure constraints using complex matrix derivatives, including diagonal structure constraints and constant modulus constraints, both of which are closely related to the state-of-the-art wireless applications. Specifically, for diagonal structure constraints mostly considered in the uplink multi-user single-input multiple-output (MU-SIMO) system and the amplitude-adjustable intelligent reflecting surface (IRS)-aided multiple-input multiple-output (MIMO) system, the capacity maximization problem, the mean-squared error (MSE) minimization problem and their variants are rigorously investigated. By leveraging complex matrix derivatives, the optimal solutions of these problems are directly obtained in closed forms. Nevertheless, for constant modulus constraints with the intrinsic nature of element-wise decomposability, which are often seen in the hybrid analog-digital MIMO system and the fully-passive IRS-aided MIMO system, we firstly explore inherent structures of the element-wise phase derivatives associated with different optimization problems. Then, we propose a novel alternating optimization (AO) algorithm with the aid of several arbitrary feasible solutions, which avoids the complicated matrix inversion and matrix factorization involved in conventional element-wise iterative algorithms. Numerical simulations reveal that the proposed algorithm can dramatically reduce the computational complexity without loss of system performance.
\end{abstract}

\vspace{-1mm}


\vspace{-3mm}
\section{Introduction}\label{Sec_intro}
Multi-antenna technology opens a new era for wireless communications due to its effective
utilization of limited spatial resources \cite{JYang_TCOM_multiante, SSugiura_CST_multiante,Telatar1999}. 
From the mathematical viewpoint, the deployment
of multi-antenna arrays at transceivers generally leads to matrix-variate optimization problems
\cite{ Xing2021, EVlachos_matrix_var,Palomar03}. 
Specifically, in the typical multiple-input multiple-output (MIMO) communication
systems, the transmit beamformer optimization and the receive equalizer optimization can be
both modeled as matrix-variate optimization problems 
\cite{HVaezy_TCOM_MIMO,ZChen_tvt_mimo,Ayach_twc_mimo}.
Compared to scalar-variate optimization, matrix-variate optimization is generally more challenging to tackle because it
inherently involves complex matrix operations, including matrix determinant, inversion, matrix
decomposition and so on. In fact, with the development of wireless communications, many
matrix-variate optimization problems with special structure constraints such as symmetric, diagonal and constant modulus structure constraints are emerging, which are closely related to the
state-of-the-art wireless systems equipped with multi-antenna transceiver antenna arrays.



In general, a structure of the matrix variable strongly depends on three factors, namely,
network architectures, frequency bands and communication demands. 
First, the distributed
network architecture has been studied, since its involved distributed antenna
arrays are capable of increasing spatial diversity gain and extending communication coverage, as
compared with the centralized counterpart \cite{WMJang_TWC_disantenna}. 
In this distributed network, the corresponding matrix variable usually has a diagonal structure. 
Second, for high-frequency millimeter wave
(mmWave) and terahertz (THz) communications, a hybrid analog-digital transceiver structure
has been regarded as an economic and effective way to achieve a large array gain \cite{SHan_hybrid,XYu_JSTSP_Hybrid}, 
in which the analog beamforming matrix is usually subject to the nonconvex and intractable constant
modulus constraints. 
Third, for smartly reconfiguring the wireless environment
in a cost-effective manner, intelligent reflecting surfaces (IRSs) composed of a large number
of passive reflecting elements have attracted a lot of attention recently \cite{TQiao_Tcom_IRS,JXu_TWC_IRS,XJin_china_IRS}. 
Considering different levels of hardware implementation, there are two main types of IRS structures that
are widely studied, i.e., the amplitude-adjustable IRS and the fully-passive IRS. Note that the
reflection matrices of these two IRSs can be mathematically modeled as diagonal matrices. In
particular, for the fully-passive IRS, the corresponding reflection matrix is additionally subject to
the nonconvex constant modulus constraint. Building upon the above discussions, it is clear that
matrix-variate optimization problems with special structure constraints have been widely considered in the state-of-the-art wireless systems. 
Therefore, it is essential to develop a framework
for optimization algorithms with guaranteed performance and low complexity for the matrix-variate
optimization.

\begin{table*}[t]
  \renewcommand\arraystretch{1.25}
  \vspace{-12mm}
	\centering
	\caption{Complex Matrix Derivatives under  Diagonal Structure Constraints}  \label{tab1}
	\vspace{-2mm}
  \resizebox{\textwidth}{!}{
		\begin{tabular}{|p{4.8cm}|p{5.2cm}|p{4.6cm}|}
			\hline
			\textbf{Function Type}  & \textbf{Derivative w.r.t. }$\bm{\Lambda}_{\bm{\Theta}}$ &  \textbf{Derivative w.r.t. }$\bm{\Lambda}_{\bm{\Theta}}^*$\\
			\hline
      $f_{\rm{D,TL}}={\rm{Tr}}(\bm{\Lambda}_{\bm{\Theta}}^{\rm{H}}\bm{M}) +{\rm{Tr}}(\bm{\Lambda}_{\bm{\Theta}}\bm{M}^{\rm{H}}) $
      & ${\rm{Diag}}\{\bm{M}^{\rm{H}}\}$ & ${\rm{Diag}}\{\bm{M}\}$\\
			\hline
      $f_{\rm{D,TQ}}={\rm{Tr}}(\bm{\Lambda}_{\bm{\Theta}}^{\rm{H}}\bm{W}\bm{\Lambda}_{\bm{\Theta}}) $
      & ${\rm{Diag}}\{\bm{\Lambda}_{\bm{\Theta}}^{\rm{H}}\bm{W}\} $   & ${\rm{Diag}}\{\bm{W} \bm{\Lambda}_{\bm{\Theta}}\}$\\
      \hline
      $ f_{\rm{D,TI}}={\rm{Tr}}\left(\left(\bm{I}_M+\bm{\Phi}\bm{\Lambda}_{\bm{\Theta}}\right)^{-1}\right)$
      &$-{\rm{Diag}}\left\{ \bm{\Phi}^{\frac{1}{2}}(\bm{I}_M+\bm{\Phi}^{\frac{1}{2}}\bm{\Lambda}_{\bm{\Theta}}\bm{\Phi}^{\frac{1}{2}})^{-2} \bm{\Phi}^{\frac{1}{2}} \right\}$& $0$\\
      \hline
      $f_{\rm{D,LD}}=\log|\bm{I}_M+\bm{\Phi}\bm{\Lambda}_{\bm{\Theta}}|$
      &${\rm{Diag}}\left\{ \bm{\Phi}^{\frac{1}{2}} (\bm{I}_M+\bm{\Phi}^{\frac{1}{2}}\bm{\Lambda}_{\bm{\Theta}}\bm{\Phi}^{\frac{1}{2}})^{-1} \bm{\Phi}^{\frac{1}{2}} \right\}$ & $0$\\
      \hline
	\end{tabular}}
  \vspace{-5mm}
\end{table*}

Currently, there have been many common popular algorithms for solving matrix-variate optimization
problems, such as the Karush-Kuhn-Tucker (KKT)-based algorithm \cite{Shiqi_TCOMKKT,Xing_KKT,S_Zhang_IRS,XinZhaoRIS}, 
the block coordinate descent (BCD) algorithm\cite{Wei_Yu_Hybrid} and the majorization-minimization (MM)-based algorithm
\cite{SGong_mm}. 
It is well-known that for convex matrix-variate problems, the KKT-based algorithm is able to
directly derive the optimal structures of matrices. Generally, the symmetric structure constraints
of matrix variables can be implicitly satisfied by the derived optimal closed-form solutions
\cite{YLi_TVT_symme}. 
Moreover, for diagonal structure constraints, this algorithm considers applying the first-order
derivative to each diagonal element to obtain the optimal solution. Furthermore, in terms of the
intractable constant modulus constraints, dual variables are usually introduced and iteratively
optimized by the subgradient method to satisfy complementary slackness conditions, thereby
potentially suffering from high iteration complexity \cite{htwAI_bcd}. 
In contrast, the BCD algorithm is
always adopted to solve highly nonconvex problems caused by strongly-coupled matrix variables.
Specifically, under the BCD framework, the original matrix-variate optimization problem can be
decomposed into multiple low-dimensional subproblems, each of which needs to be iteratively
optimized until convergence. In order to ensure at least local convergence of the BCD algorithm
\cite{Wei_Yu_Hybrid}, each subproblem is required to have a unique optimal solution. Nevertheless, considering
that subproblems may be nonconvex and thus hard to globally solve, the MM-based algorithm
has attained extensive attention, whose core idea is to construct a tractable surrogate function
to locally approximate the original nonconvex subproblem. Unfortunately, the derivation of the
surrogate function usually involves high-complexity matrix manipulations and also needs to be
iteratively carried out to achieve a close approximation.

Obviously, the KKT-based algorithm based on complex matrix derivatives generally achieves
the lowest complexity among the three types of algorithms\cite{AHj_Complex_Matrix_Derivatives}. Nonetheless, its application range is
relatively limited as compared to the BCD and MM-based algorithms. Note that the implementation of the latter two algorithms depends on the specific wireless system and may have high
computational complexity, especially for large-scale arrays. To circumvent these issues, in this
paper, we aim to develop a unified framework for matrix-variate optimization with two special
structure constraints, namely, diagonal structure and constant modulus constraints. For each
considered case, the novel low-complexity algorithm with guaranteed performance is proposed.
The main contributions of our work are further summarized as follows.
\begin{itemize}
  \vspace{-1.5mm}
\item Firstly, we consider the diagonal structure constraints often seen in the uplink multi-user single-input multiple-output (MU-SIMO) system and the amplitude-adjustable IRS-aided MIMO system, which
are always involved in the capacity maximization problem, mean squared error (MSE)
minimization problem and their variants. We propose complex matrix derivatives associated
with diagonal structures, based on which the optimal solutions of these matrix-variate
problems are directly obtained in closed forms. Furthermore, the above study is extended
to the case of block-diagonal structure constraints.

\item Secondly, in terms of constant modulus constraints mostly adopted in the hybrid analog-digital MIMO system and the fully-passive IRS-aided MIMO system, we propose the
element-wise phase derivatives inspired by their element-wise decomposability nature. For
different classical matrix-variate optimization problems, it is revealed that the element-wise
phase derivatives can be classified into the following two general forms, i.e., the linear form
and the conjugate linear form.

\item Finally, by exploring inherent structures of the element-wise phase derivatives, we develop
a novel alternating optimization (AO) algorithm with the aid of several arbitrary feasible
solutions for the matrix-variate optimization under constant modulus constraints. Note
that the computational complexity of the proposed AO algorithm sharply decreases, since it
avoids the complicated matrix inversion and matrix factorization involved in the conventional
element-wise iterative algorithm. Moreover, we demonstrate that the proposed algorithm is
able to achieve almost the same performance as the existing benchmark schemes.

\vspace{-1mm}
\end{itemize}


\textit{Notation:}
Scalars, vectors and matrices are represented by non-bold, bold lowercase, and bold uppercase letters, respectively.
The notations $\bm{A}^{\rm{T}}, \bm{A}^*, \bm{A}^{\rm{H}}, \bm{A}^{-1}$, ${\rm{Tr}}(\bm{A})$ and $\vert \bm{A} \vert $ denote the transpose, conjugate, hermitian, inversion, trace and determinant of the complex matrix $\bm{A}$, respectively.
${\rm{Diag}}\{\bm{A}\}$ denotes a vector whose elements are diagonal elements of matrix $\bm{A}$, 
and ${\rm{Blockdiag}}(\{\bm{A}_k\}_{k=1}^K)$ is a block diagonal matrix with diagonal sub-matrices of $\bm{A}_k$'s.
Moreover, the $i$th row and the $j$th column of $\bm{A}$ are denoted as $[\bm{A}]_{i,:}$ and $[\bm{A}]_{:,j}$, respectively, 
the element in the $i$th row and the $j$th column is denoted as $[\bm{A}]_{i,j}$.
$\frac{{\rm{d}} f}{{\rm{d}} a}$ and $\frac{\partial f }{\partial a}$ denote the differential and the partial derivative of $f$ with respect to $a$, respectively.
$\odot$ denotes the Hadamard product and $(a)^+ = \max\{0, a\}$.
$\Re\{a\}$ and $\Im\{a\} $ denote the real and imaginary parts of a complex variable $a$, respectively.
Lastly, the word ``with respect to'' is abbreviated as ``w.r.t.''.


\vspace{-2.5mm}
\section{Diagonal Structure Constraints}\label{sec_diag}

\begin{figure*}[t]
  \vspace{-12mm}
    \centering
    \includegraphics[width=0.9\textwidth]{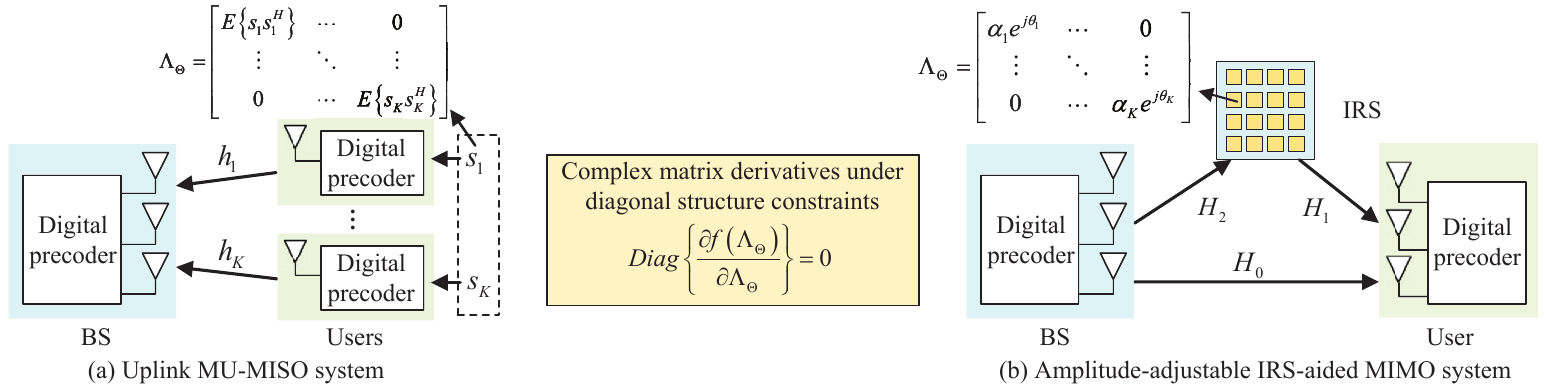}
	\vspace{-3mm}
    \caption{A  diagram of  application scenarios associated  with diagonal matrix variables.}
   \label{fig_sys_diag}
    \vspace{-5mm}
\end{figure*}

In this section, we firstly provide some fundamental properties of complex matrix derivatives associated with diagonal structures for several types of objective functions.
Based on these properties,  we  then obtain the optimal solutions of a series of   optimization problems in the uplink MU-SIMO system and the amplitude-adjustable IRS-aided MIMO system  in closed forms.
Moreover, the above study is extended to the case of block-diagonal matrix variables.

\vspace{-3.5mm}
\subsection{Mathematical Preliminaries}\label{sec_diag_math}

At the beginning, some fundamental definitions for diagonal matrices are provided, which are the basis of the following analysis. 
For any two diagonal matrices $\bm{\Lambda}_{\bm{\Theta}_1} \in \mathbb{C}^{M \times M} $ and $\bm{\Lambda}_{\bm{\Theta}_2}\in \mathbb{C}^{M \times M}$, we have
\begin{align}
{\rm{Diag}}\{\bm{\Lambda}_{\bm{\Theta}_1}+\bm{\Lambda}_{\bm{\Theta}_2}\}
=& \ {\rm{Diag}}\{\bm{\Lambda}_{\bm{\Theta}_1}\}
+{\rm{Diag}}\{\bm{\Lambda}_{\bm{\Theta}_2}\}.
\end{align}
For an arbitrary square matrix $\bm{\Phi}\in \mathbb{C}^{M \times M}$, we have 
\begin{align}
{\rm{Diag}}\{\bm{\Phi}\bm{\Lambda}_{\bm{\Theta}} \}={\rm{Diag}}\{\bm{\Lambda}_{\bm{\Theta}}\bm{\Phi} \}.
\end{align}
Moreover, the following equality holds for any complex matrices $\bm{N}\in \mathbb{C}^{N \times M}$ and $\bm{M}\in \mathbb{C}^{M \times N}$, i.e., 
\begin{align}\label{diag_prop_MN}
{\rm{Diag}}\{\bm{N}\bm{\Lambda}_{\bm{\Theta}}\bm{M}\}=\bm{\Phi} {\rm{Diag}}\{\bm{\Lambda}_{\bm{\Theta}}\} \ \text{with}\ \bm{\Phi}=\bm{N} \odot \bm{M}^{\rm{T}}.
\end{align}
Together with the operation of ${\rm{Diag}}\{\cdot\}$, the vectorization of the first-order derivative of a scalar-valued function $f(\bm{\Lambda}_{\bm{\Theta}})$ w.r.t the complex diagonal matrix $\bm{\Lambda}_{\bm{\Theta}}$ can be defined as
\cite{AHj_Complex_Matrix_Derivatives}
\begin{align}
& {\rm{Diag}}\left\{\frac{\partial f(\bm{\Lambda}_{\bm{\Theta}})}{\partial \bm{\Lambda}_{\bm{\Theta}}}\right\}= \left[ \frac{\partial f(\bm{\Lambda}_{\bm{\Theta}})}{\partial [\bm{\Lambda}_{\bm{\Theta}}]_{1,1}},\cdots,\frac{f(\bm{\Lambda}_{\bm{\Theta}})}{\partial [\bm{\Lambda}_{\bm{\Theta}}]_{N,N}} \right]^{\rm{T}}, \nonumber \\
& {\rm{Diag}}\left\{\frac{\partial f(\bm{\Lambda}_{\bm{\Theta}})}{\partial \bm{\Lambda}_{\bm{\Theta}}^*}\right\}= \left[ \frac{\partial f(\bm{\Lambda}_{\bm{\Theta}})}{\partial [\bm{\Lambda}_{\bm{\Theta}}]_{1,1}^*},\cdots,\frac{f(\bm{\Lambda}_{\bm{\Theta}})}{\partial [\bm{\Lambda}_{\bm{\Theta}}]_{N,N}^*} \right]^{\rm{T}}.
\end{align} 

In the sequel, we mainly concern about complex matrix derivatives w.r.t. diagonal matrices for four common objective functions, 
including the trace-linear function, the trace-quadratic function, the trace-inverse function and the log-determinant function.

\subsubsection{ Trace-Linear Function}
For an arbitrary complex matrix $\bm{M}\!\!\in \!\!\mathbb{C}^{M\! \times \!M}$, the differential of a trace-linear function 
$f_{\rm{D,\!TL}}\!\!=\!{\rm{Tr}}(\bm{\Lambda}_{\bm{\Theta}}^{\rm{H}}\bm{M})\! +\!{\rm{Tr}}(\bm{\Lambda}_{\bm{\Theta}}\bm{M}^{\rm{H}})$ w.r.t. $\bm{\Lambda}_{\bm{\Theta}}$ can be obtained as
${\rm{d}}\left(f_{\rm{D,TL}}\right)
=
{\rm{Tr}}\left({\rm{d}}
\left(\bm{\Lambda}_{\bm{\Theta}}\right) \bm{M}^{\rm{H}} \right).
$
Based on this, the corresponding first-order derivative w.r.t. $\bm{\Lambda}_{\bm{\Theta}}$ can be obtained as
\begin{align}
  {\rm{Diag}}\left\{\frac{\partial f_{\rm{D,TL}}}{\partial \bm{\Lambda}_{\bm{\Theta}}}\right\} 
  ={\rm{Diag}}\{\bm{M}^{\rm{H}}\}.
\end{align}
The first-order derivative of $f_{\rm{D,TL}}$ w.r.t. $\bm{\Lambda}_{\bm{\Theta}}^*$ is also given by
\begin{align}
  {\rm{Diag}}\left\{\frac{\partial f_{\rm{D,TL}}}{\partial \bm{\Lambda}_{\bm{\Theta}}^*}\right\} 
={\rm{Diag}}\{\bm{M}\}.
\end{align}

\subsubsection{ Trace-Quadratic Function}
For a Hermitian matrix $\bm{W}\!\!\!\in \!\!\mathbb{C}^{M \!\times\! M}$, 
the differential of a trace-quadratic function $f_{\rm{D,TQ}}\!\!=\!{\rm{Tr}}(\bm{\Lambda}_{\bm{\Theta}}^{\rm{H}}\bm{W}\bm{\Lambda}_{\bm{\Theta}}\!)$ 
w.r.t. $\bm{\Lambda}_{\bm{\Theta}}$ and $\bm{\Lambda}_{\bm{\Theta}}^*$ are respectively calculated as
 $ {\rm{d}}\left( f_{\rm{D,TQ}} \right)
  ={\rm{Tr}}(\bm{\Lambda}_{\bm{\Theta}}^{\rm{H}}\bm{W}{\rm{d}} (\bm{\Lambda}_{\bm{\Theta}})),
{\rm{d}}\left( f_{\rm{D,TQ}} \right)
={\rm{Tr}}({\rm{d}} (\bm{\Lambda}_{\bm{\Theta}}^{\rm{H}}) \bm{W} \bm{\Lambda}_{\bm{\Theta}}).
$
Then, the corresponding first-order derivatives w.r.t. $\bm{\Lambda}_{\bm{\Theta}}$ and $\bm{\Lambda}_{\bm{\Theta}}^*$ can be obtained as
\begin{align}\label{diag_pre_tq}
& {\rm{Diag}}\left\{\frac{\partial f_{\rm{D,TQ}} }{\partial \bm{\Lambda}_{\bm{\Theta}}}\right\} ={\rm{Diag}}\{\bm{\Lambda}_{\bm{\Theta}}^{\rm{H}}\bm{W}\},\nonumber\\
 &{\rm{Diag}}\left\{\frac{\partial f_{\rm{D,TQ}} }{\partial \bm{\Lambda}_{\bm{\Theta}}^*} \right\}={\rm{Diag}}\{\bm{W}\bm{\Lambda}_{\bm{\Theta}}\}.
\end{align}

\subsubsection{ Trace-Inverse Function}
For a positive semi-definite matrix $\bm{\Phi}\in \mathbb{C}^{M \times M}$, the differential  of a trace-inverse function $f_{\rm{D,TI}}={\rm{Tr}}\left(\left(\bm{I}_M+\bm{\Phi}\bm{\Lambda}_{\bm{\Theta}}\right)^{-1}\right)$ w.r.t. $\bm{\Lambda}_{\bm{\Theta}}$ is given by
${\rm{d}}\left( f_{\rm{D,TI}}
\right)=-{\rm{Tr}}\left( (\bm{I}_M+\bm{\Phi}\bm{\Lambda}_{\bm{\Theta}})^{-2}\bm{\Phi} {\rm{d}} \bm{\Lambda}_{\bm{\Theta}}\right).
$
The corresponding first-order derivative w.r.t. $\bm{\Lambda}_{\bm{\Theta}}$ is then given by
\begin{align}\label{deri_diag_tr}
  {\rm{Diag}}\left\{\frac{\partial f_{\rm{D,TI}} }{\partial \bm{\Lambda}_{\bm{\Theta}}}\right\}
=&-{\rm{Diag}}\left\{ (\bm{I}_M+\bm{\Phi}\bm{\Lambda}_{\bm{\Theta}})^{-2} \bm{\Phi} \right\}\\
=&-{\rm{Diag}}\left\{ \bm{\Phi}^{\frac{1}{2}}(\bm{I}_M+\bm{\Phi}^{\frac{1}{2}}\bm{\Lambda}_{\bm{\Theta}}\bm{\Phi}^{\frac{1}{2}})^{-2} \bm{\Phi}^{\frac{1}{2}} \right\}.\nonumber
\end{align}

\subsubsection{ Log-Determinant Function}
Considering a log-determinant function $f_{\rm{D,LD}}\!=\!\log|\bm{I}_M\!+\!\bm{\Phi}\bm{\Lambda}_{\bm{\Theta}}|$, its differential  w.r.t. $\bm{\Lambda}_{\bm{\Theta}}$ can be derived as 
  ${\rm{d}}\left(f_{\rm{D,LD}} \right)\!=\!{\rm{Tr}}\left( (\bm{I}_M\!+\!\bm{\Phi}\bm{\Lambda}_{\bm{\Theta}})^{-1}\bm{\Phi} {\rm{d}} \bm{\Lambda}_{\bm{\Theta}}\right),
$
based on which the following
first-order derivative w.r.t. $\bm{\Lambda}_{\bm{\Theta}}$ holds.
\begin{align}\label{diag_log_prop}
  {\rm{Diag}}\left\{\frac{\partial f_{\rm{D,LD}}}{\partial \bm{\Lambda}_{\bm{\Theta}}}\right\}
  =&{\rm{Diag}}\left\{ (\bm{I}_M+\bm{\Phi}\bm{\Lambda}_{\bm{\Theta}})^{-1} \bm{\Phi} \right\}\\
  =&{\rm{Diag}}\left\{ \bm{\Phi}^{\frac{1}{2}} (\bm{I}_M+\bm{\Phi}^{\frac{1}{2}}\bm{\Lambda}_{\bm{\Theta}}\bm{\Phi}^{\frac{1}{2}})^{-1} \bm{\Phi}^{\frac{1}{2}} \right\}.\nonumber
\end{align} 

In conclusion, the complex matrix derivatives of the above four types of objective functions are summarized in Table \ref{tab1}.
Exploiting these fundamental properties, some classical wireless applications are investigated in the following subsection.

\vspace{-4mm}
\subsection{Specific Wireless Applications}\label{sec_diag_app}

\subsubsection{Uplink MU-SIMO System}\label{sec_diag_hyb}
As shown in Fig. \ref{fig_sys_diag}, we firstly consider an uplink distributed MU-SIMO system, where $K$ single-antenna users transmit independent data streams to the BS
\cite{WMJang_TWC_disantenna}.
The received signal at the BS can be expressed as 
\begin{align}
\bm{y}=\bm{H}\bm{s}+\bm{n} \ \ \text{with} \ \ \bm{H}=[\bm{h}_1,\cdots,\bm{h}_K]\in \mathbb{C}^{N_{\rm{t}} \times K},
\end{align}
where $\bm{h}_k\!\in \!\mathbb{C}^{N_{\rm{t}} \times 1}$ denotes the channel between the BS and the $k$th user, and $\bm{s}\!=\![ s_1,\cdots,s_K ]^{\rm{T}}\!\!\!\in\! \mathbb{C}^{K \times 1}$ is the transmit signal, 
$\bm{n}$ is the additive noise obeying Gaussian distribution with zero mean and covariance matrix $\mathbb{E}\{\bm{n}\bm{n}^{\rm{H}}\}\!=\!\bm{\Sigma}$.
It is worth noting that the covariance matrix of $\bm{s}$ is diagonal, 
since the transmit data streams of $K$ users are independent of each other, i.e.,
$\mathbb{E}\{\bm{s}{\bm{s}}^{\rm{H}}\}=\bm{\Lambda}_{\bm{\Theta}}$.
Then, the uplink capacity maximization problem is formulated as
\begin{align}
\textbf{Prob.1:} \ \max_{\bm{\Lambda}_{\bm{\Theta}}} \ & \log\left|\bm{I}_{N_t}+\bm{\Sigma}^{-1}\bm{H}\bm{\Lambda}_{\bm{\Theta}}\bm{H}^{\rm{H}}\right| \nonumber \\
{\rm{s.t.}} \ &  {\rm{Tr}}(\bm{\Lambda}_{\bm{\Theta}})\le P, \
 0\leq[\bm{\Lambda}_{\bm{\Theta}}]_{k,k}\leq P_k, \ \forall k.
\end{align}
By recalling (\ref{diag_log_prop}), the first-order derivative of the objective function of \textbf{Prob.1} w.r.t. $\bm{\Lambda}_{\bm{\Theta}}$ is given by
\begin{align}
  &{\rm{Diag}}\left\{ \frac{\partial \log|\bm{I}_{N_t}+\bm{\Sigma}^{-1}\bm{H}\bm{\Lambda}_{\bm{\Theta}}\bm{H}^{\rm{H}}|}{\partial \bm{\Lambda}_{\bm{\Theta}}}\right\}\\
  =&{\rm{Diag}}\!\left\{\bm{H}^{\rm{H}}\bm{\Sigma}^{-\frac{1}{2}}
\left(\bm{I}_{N_t}\!+\!\bm{\Sigma}^{-\frac{1}{2}}\bm{H}\bm{\Lambda}_{\bm{\Theta}}\bm{H}^{\rm{H}}\bm{\Sigma}^{-\frac{1}{2}}\right)^{-1}
\bm{\Sigma}^{-\frac{1}{2}}\bm{H}\right\}.\nonumber
\end{align} 
In order to derive the optimal $\bm{\Lambda}_{\bm{\Theta}}$, 
we present the KKT optimality conditions of \textbf{Prob.1} as follows
\cite{Boyd04}:
  \begin{subequations}
    \begin{numcases}{}
{\rm{Diag}}\bigg\{\bm{H}^{\rm{H}}\bm{\Sigma}^{-\frac{1}{2}}
\left(\bm{I}_{N_t}+\bm{\Sigma}^{-\frac{1}{2}}\bm{H}\bm{\Lambda}_{\bm{\Theta}}\bm{H}^{\rm{H}}
\bm{\Sigma}^{-\frac{1}{2}}\right)^{-1}  \nonumber\\
 \quad \quad \  \times \bm{\Sigma}^{-\frac{1}{2}}\bm{H}\bigg\}
=\mu {\rm{Diag}}\left\{\bm{I}_K\right\} \!-\! [\psi_1,\cdots,\psi_K]^{\rm{T}},\label{Prob_2_Condition_a}\\
\mu \left( {\rm{Tr}}(\bm{\Lambda}_{\bm{\Theta}})- P\right)=0,\label{Prob_2_Condition_b}\\
\psi_k \left([\bm{\Lambda}_{\bm{\Theta}}]_{k,k}- P_k \right)\! =\!0,\ \forall k,
  \end{numcases}
     \end{subequations} 
where $\mu$ is the Lagrange multiplier associated with the sum power constraint and $\psi_k$'s correspond to the power constraints imposed on each user.
We firstly define 
$  \bm{J}_k=\bm{I}_K+\left(\bm{H}^{\rm{H}}\bm{\Sigma}^{-1}\bm{H} - \overline{\bm H}_k\right)  
  \left(\bm \Lambda_{\bm{\Theta}}-\overline{ \bm \Lambda_{\bm{\Theta}}}_{k}\right) 
$,
where 
$\overline{\bm H}_k\in \mathbb{C}^{K \times K}$ is an all-zero matrix except for its $k$th column being $\left[\bm{H}^{\rm{H}}\bm{\Sigma}^{-1}\bm{H}\right]_{:,k}$
and $\overline{ \bm \Lambda_{\bm{\Theta}}}_{k}\in \mathbb{C}^{K \times K}$ is an all-zero matrix except for its $(k,k)$th element being $\left[\bm \Lambda_{\bm{\Theta}}\right]_{k,k}$.
Then, the left-hand side of (\ref{Prob_2_Condition_a}) can be rewritten as
\begin{align}\label{left_diag_log}
  &{\rm{Diag}}\left\{\bm{H}^{\rm{H}}\bm{\Sigma}^{-\frac{1}{2}}
\!\left(\bm{I}_{N_t}\!\!+\!\!\bm{\Sigma}^{-\frac{1}{2}}\bm{H}\bm{\Lambda}_{\bm{\Theta}}\bm{H}^{\rm{H}}
\bm{\Sigma}^{-\frac{1}{2}}\right)^{-1}\!
\bm{\Sigma}^{-\frac{1}{2}}\bm{H}\right\}\nonumber\\
\overset{(a_1)}{=}&{\rm{Diag}}\left\{
\left(\bm{I}_K+\bm{H}^{\rm{H}}\bm{\Sigma}^{-1}\bm{H}\bm{\Lambda}_{\bm{\Theta}}
\right)^{-1}
\bm{H}^{\rm{H}}\bm{\Sigma}^{-1}\bm{H}\right\}\nonumber\\
\overset{(a_2)}{=}& 
-\frac{[\bm{\Lambda}_{\bm{\Theta}}]_{k,k}{\rm{Diag}}\left\{\bm{J}_k^{-1} \overline{\bm H}_k \bm{J}_k^{-1} \bm{H}^{\rm{H}}\bm{\Sigma}^{-1}\bm{H}\right\}}{1+[\bm{\Lambda}_{\bm{\Theta}}]_{k,k} {\rm{Tr}} (\bm{J}_k^{-1} \overline{\bm H}_k)}  
\nonumber\\
&+{\rm{Diag}}\left\{\bm{J}_k^{-1} \bm{H}^{\rm{H}}\bm{\Sigma}^{-1}\bm{H}\right\}, 
\end{align}
where $(a_1)$ holds based on the matrix inversion lemma $\bm{N}(\bm{I}+\bm{M}\bm{N})^{-1}=(\bm{I}+\bm{N}\bm{M})^{-1}\bm{N}$
and $(a_2)$ is attained using the Sherman Morrison formula, i.e., $(\bm{M}+\bm{N})^{-1}=\bm{M}^{-1}-\frac{\bm{M}^{-1} \bm{N} \bm{M}^{-1}}{1+{\rm{Tr}}\left(\bm{M}^{-1} \bm{N}\right)}$ for a full-rank matrix $\bm{M}$ and a rank-one matrix $\bm{N}$.
By substituting (\ref{left_diag_log}) into (\ref{Prob_2_Condition_a}), the optimal $[\bm{\Lambda}_{\bm{\Theta}}]_{k,k}$'s is derived in the following water-filling form \cite{YYu_TSP_WF}.
\begin{align}\label{opt_hy_diag_log}
  [\bm{\Lambda}_{\bm{\Theta}}]_{k,k}\!\!=\!\!\!
  \begin{cases}
	\!\!\!\!\left( \!\! \frac{y_k} {\left[ {\rm{Diag}}\left\{\!\bm{J}_k^{-1} \overline{\bm H}_k \bm{J}_k^{-1} \!\bm{H}^{\rm{H}}\!\bm{\Sigma}^{-1}\!\!\bm{H}\!\right\}\!\right]_k 
  \! \!+\! {\rm{Tr}} (\!\bm{J}_k^{-1} \overline{\bm H}_k\!) y_k }
  \!\!\right)^{+}\!\!\!\!,\\
 \qquad \ \ \! \text { if } [\bm{\Lambda}_{\bm{\Theta}}]_{k,k}\leq P_k,  \\
		P_k, \quad \text { if } [\bm{\Lambda}_{\bm{\Theta}}]_{k,k} > P_k, \ \forall k ,
		\end{cases}
\end{align}
where $y_k=\left[ {\rm{Diag}}\left\{\bm{J}_k^{-1} \bm{H}^{\rm{H}}\bm{\Sigma}^{-1}\bm{H}\right\}\right]_k -\mu$. 
Moreover, since ${\rm{Tr}}(\bm{\Lambda}_{\bm{\Theta}})$ is monotonically decreasing w.r.t $\mu>0$, the optimal $\mu$ satisfying (\ref{Prob_2_Condition_b}) can be found via the
bisection search.

In addition, MSE is a widely used performance metric, which reflects the accuracy of the desired signals that can be recovered from the noise corrupted observations.  
Accordingly, the MSE minimization problem is formulated as 
\begin{align}
\textbf{Prob.2:} \ \min_{\bm{\Lambda}_{\bm{\Theta}}} \ & {\rm{Tr}}\left[\left(\bm{I}_{N_t}+\bm{\Sigma}^{-1}\bm{H}\bm{\Lambda}_{\bm{\Theta}}\bm{H}^{\rm{H}}\right)^{-1} \right] \nonumber \\
{\rm{s.t.}} \ & {\rm{Tr}}(\bm{\Lambda}_{\bm{\Theta}})\le P, \ 0\leq [\bm{\Lambda}_{\bm{\Theta}}]_{k,k}\leq P_k, \ \forall k.
\end{align}
By recalling (\ref{deri_diag_tr}), the first-order derivative of the objective function of \textbf{Prob.2} w.r.t. $\bm{\Lambda}_{\bm{\Theta}}$ is given by
\begin{align}\label{Prob_3_Derivative}
&{\rm{Diag}}\left\{\frac{\partial {\rm{Tr}}\left[(\bm{I}_{N_t}+\bm{\Sigma}^{-1}\bm{H}\bm{\Lambda}_{\bm{\Theta}}\bm{H}^{\rm{H}})^{-1}\right]}{\partial \bm{\Lambda}_{\bm{\Theta}}}\right\}\\
=&\!-\! {\rm{Diag}}\!\left\{\bm{H}^{\rm{H}}\bm{\Sigma}^{-\frac{1}{2}}
\!\!\left(\bm{I}_{N_t}\!+\!\bm{\Sigma}^{-\frac{1}{2}}\bm{H}\bm{\Lambda}_{\bm{\Theta}}\bm{H}^{\rm{H}}
\bm{\Sigma}^{-\frac{1}{2}}\right)^{-2}\!\!
\bm{\Sigma}^{-\frac{1}{2}}\bm{H}\right\}\!.\nonumber
\end{align}
Based on (\ref{Prob_3_Derivative}), the KKT optimality conditions of \textbf{Prob.2} can be formulated as
  \begin{subequations}
    \begin{numcases}{}
   {\rm{Diag}}\left\{\bm{H}^{\rm{H}}\bm{\Sigma}^{-\frac{1}{2}}
\left(\bm{I}_{N_t}+\bm{\Sigma}^{-\frac{1}{2}}\bm{H}\bm{\Lambda}_{\bm{\Theta}}\bm{H}^{\rm{H}}
\bm{\Sigma}^{-\frac{1}{2}}\right)^{-2}
\right. \nonumber\\
\left. \quad \quad \  \times \bm{\Sigma}^{-\frac{1}{2}}\bm{H}\right\}
=\mu  {\rm{Diag}}\left\{\bm{I}_K\right\} \!-\![\psi_1,\cdots,\psi_K]^{\rm{T}},\label{Prob_4_Condition_a}\\
   \mu \left( {\rm{Tr}}(\bm{\Lambda}_{\bm{\Theta}})- P\right)=0,\label{Prob_4_Condition_b}\\
  \psi_k \left([\bm{\Lambda}_{\bm{\Theta}}]_{k,k}- P_k \right) \!=\!0,\ \forall k,
    \end{numcases} 
  \end{subequations}  
where $\mu$ and $\psi_k$'s are defined similarly to \textbf{Prob.1}.
The left-hand side of (\ref{Prob_4_Condition_a}) can be rewritten as
\begin{align}\label{left_diag_tr}
  &{\rm{Diag}}\!\left\{\bm{H}^{\rm{H}}\bm{\Sigma}^{-\frac{1}{2}}
\left(\bm{I}_{N_t}\!+\!\bm{\Sigma}^{-\frac{1}{2}}\bm{H}\bm{\Lambda}_{\bm{\Theta}}\bm{H}^{\rm{H}}
\bm{\Sigma}^{-\frac{1}{2}}\right)^{-2}
\bm{\Sigma}^{-\frac{1}{2}}\bm{H}\right\}\nonumber\\
\overset{(b)}{=}&{\rm{Diag}}\left\{ 
  \left(\bm{J}_k^{-1} 
-\frac{[\bm{\Lambda}_{\bm{\Theta}}]_{k,k}\bm{J}_k^{-1} \overline{\bm H}_k \bm{J}_k^{-1} }{1+[\bm{\Lambda}_{\bm{\Theta}}]_{k,k} {\rm{Tr}} (\bm{J}_k^{-1} \overline{\bm H}_k)}  
\right)^{2}\bm{H}^{\rm{H}}\bm{\Sigma}^{-1}\bm{H}\right\} \nonumber\\
=& \frac{[\bm{\Lambda}_{\bm{\Theta}}]_{k,k}^2 {\rm{Diag}}\left\{\bm{J}_k^{-1} \overline{\bm H}_k \bm{J}_k^{-2}\overline{\bm H}_k \bm{J}_k^{-1}  \bm{H}^{\rm{H}}\bm{\Sigma}^{-1}\bm{H} \right\}}{\left( 1+[\bm{\Lambda}_{\bm{\Theta}}]_{k,k} {\rm{Tr}} (\bm{J}_k^{-1} \overline{\bm H}_k)\right)^2} \nonumber\\
&-\frac{[\bm{\Lambda}_{\bm{\Theta}}]_{k,k} {\rm{Diag}}\left\{  \bm{J}_k^{-2} \overline{\bm H}_k \bm{J}_k^{-1} \bm{H}^{\rm{H}}\bm{\Sigma}^{-1}\bm{H}\right\}}{1+[\bm{\Lambda}_{\bm{\Theta}}]_{k,k} {\rm{Tr}} (\bm{J}_k^{-1} \overline{\bm H}_k)}\nonumber\\
&-\frac{[\bm{\Lambda}_{\bm{\Theta}}]_{k,k} {\rm{Diag}}\left\{  \bm{J}_k^{-1} \overline{\bm H}_k \bm{J}_k^{-2} \bm{H}^{\rm{H}}\bm{\Sigma}^{-1}\bm{H}\right\}}{1+[\bm{\Lambda}_{\bm{\Theta}}]_{k,k} {\rm{Tr}} (\bm{J}_k^{-1} \overline{\bm H}_k)} \nonumber\\
&+{\rm{Diag}}\left\{\bm{J}_k^{-2} \bm{H}^{\rm{H}}\bm{\Sigma}^{-1}\bm{H}
\right\},
\end{align}
where $(b)$ holds similarly to (\ref{left_diag_log}).
Substituting (\ref{left_diag_tr}) into (\ref{Prob_4_Condition_a}), we have
\begin{align}
  \widetilde{a}_k [\bm{\Lambda}_{\bm{\Theta}}]_{k,k}^2 
  +\widetilde{b}_k [\bm{\Lambda}_{\bm{\Theta}}]_{k,k}
  +x_k=0, \ \forall k,
\end{align}
where 
\begin{align}
  \widetilde{a}_k=& {\rm{Tr}}^2 (\bm{J}_k^{-1} \overline{\bm H}_k)x_k
  \!+\!\big[{\rm{Diag}}\left\{\bm{J}_k^{-1} \overline{\bm H}_k \bm{J}_k^{-2}\overline{\bm H}_k \bm{J}_k^{-1}\!  \bm{H}^{\rm{H}}
  \bm{\Sigma}^{-1}\right. \nonumber\\
&\left.\!\bm{H} \right\}\big]_k \!-\! {\rm{Tr}} (\bm{J}_k^{-1} \overline{\bm H}_k)
[{\rm{Diag}}\left\{  \bm{J}_k^{-2} \overline{\bm H}_k \bm{J}_k^{-1}\!
 \bm{H}^{\rm{H}}\bm{\Sigma}^{-1}\!\bm{H}\right\}]_k\nonumber\\
&- {\rm{Tr}} (\bm{J}_k^{-1} \overline{\bm H}_k)
[{\rm{Diag}}\left\{ \bm{J}_k^{-1} \overline{\bm H}_k \bm{J}_k^{-2} \bm{H}^{\rm{H}}\bm{\Sigma}^{-1}\bm{H}\right\}]_k,\nonumber\\
\widetilde{b}_k=&{2 \rm{Tr}} (\bm{J}_k^{-1} \overline{\bm H}_k)x_k 
-[{\rm{Diag}}\left\{  \bm{J}_k^{-2} \overline{\bm H}_k \bm{J}_k^{-1}
\bm{H}^{\rm{H}}\bm{\Sigma}^{-1}\bm{H}\right\}]_k\nonumber\\
&-[{\rm{Diag}}\left\{ \bm{J}_k^{-1} \overline{\bm H}_k \bm{J}_k^{-2} \bm{H}^{\rm{H}}\bm{\Sigma}^{-1}\bm{H}\right\}]_k,\nonumber\\
x_k=&\left[{\rm{Diag}}\left\{\bm{J}_k^{-2} \bm{H}^{\rm{H}}\bm{\Sigma}^{-1}\bm{H}\right\}\right]_k  -\mu, \ \forall k .
\end{align}
Then, the optimal $[\bm{\Lambda}_{\bm{\Theta}}]_{k,k}$'s of \textbf{Prob.2} can be attained using quadratic formula as follows:
\begin{align}\label{opt_hy_diag_tr}
  [\bm{\Lambda}_{\bm{\Theta}}]_{k,k}\!=\!\!
  \begin{cases}
    \!\!\!	\left( \!\! \frac{ -\widetilde{b} \pm \sqrt{ \widetilde{b}^2- 4 [\bm{x}_k]_k\widetilde{a}}}{2 \widetilde{a}}
    \!\right)^{+}
\!\!\!, \! \!\!\!\!\!&\text { if } [\bm{\Lambda}_{\bm{\Theta}}]_{k,k}\leq P_k,\\
		P_k, \! \!\!\!\!\!&\text { if } [\bm{\Lambda}_{\bm{\Theta}}]_{k,k} > P_k,
		\end{cases} \ \forall k.
\end{align}
Similarly, $\mu$  can be obtained using
the bisection search. 

\subsubsection{Amplitude-Adjustable IRS-aided MIMO System}
Hereafter, we consider the state-of-the-art amplitude-adjustable IRS-aided point-to-point MIMO system as shown in Fig. \ref{fig_sys_diag},
in which the received signal at the user can be expressed as 
\begin{align}\label{diag_irs_channel}
\bm{y}=\bm{H}\bm{s}+\bm{n} \ \ \text{with} \ \ \bm{H}=\bm{H}_0+\bm{H}_1\bm{\Lambda}_{\bm{\Theta}}\bm{H}_2,
\end{align}
where $\bm{H}_0\!\in\! \mathbb{C}^{N_r\! \times\! N_t}$, $\bm{H}_1\!\in\! \mathbb{C}^{N_r\! \times\! K}$ and $\bm{H}_2\!\!\in \!\!\mathbb{C}^{K \times N_t}$ represent the BS-user direct channel, the IRS-user channel and the BS-IRS channel, respectively.
$\bm{\Lambda}_{\bm{\Theta}}\!\in\! \mathbb{C}^{K \!\times\! K}$ denotes the diagonal IRS reflection matrix whose each diagonal element represents the adjustable amplitude and phase of the corresponding reflecting element, 
which usually satisfies $\left\lvert \left[\bm{\Lambda}_{\bm{\Theta}}\right]_{i,i} \right\rvert \!\leq\! 1$
or ${\rm{Tr}}(\bm{\Lambda}_{\bm{\Theta}}\bm{\Lambda}_{\bm{\Theta}}^{\rm{H}}) \!\le\! K$ \cite{SAbey_Tcom_irsamp}.
Similar to Sec. \ref{sec_diag_hyb}, we firstly consider
the following capacity maximization problem.
\begin{align}
\textbf{Prob.3:} \ \max_{\bm{\Lambda}_{\bm{\Theta}}} \ 
& \log \left|\bm{\Sigma}^{-1} 
\bm{H} \bm{H}^{\rm{H}}+\bm{I}_{N_r}\right|\nonumber \\
{\rm{s.t.}} \ & {\rm{Tr}}(\bm{\Lambda}_{\bm{\Theta}}\bm{\Lambda}_{\bm{\Theta}}^{\rm{H}}) \le K.
\end{align} 
By recalling (\ref{diag_pre_tq}) and (\ref{diag_log_prop}), the first-order derivative of the objective function of \textbf{Prob.3} w.r.t. $\bm{\Lambda}_{\bm{\Theta}}$ is given by
\begin{align}\label{diag_irs_log_deri}
& {\rm{Diag}}\left\{\frac{\partial \log \left|\bm{\Sigma}^{-1}
\bm{H} \bm{H}^{\rm{H}}
+\bm{I}_{N_r}\right|
}{\partial \bm{\Lambda}_{\bm{\Theta}}}\right\} \nonumber \\
=&{\rm{Diag}}\left\{  \bm{H}_2 \bm{H}_2^{\rm{H}} \bm{\Lambda}_{\bm{\Theta}}^{\rm{H}} \bm{H}_1^{\rm{H}} \left[\bm{\Sigma}^{-1}
\bm{H} \bm{H}^{\rm{H}}
+\bm{I}_{N_r}\right]^{-1} \bm{\Sigma}^{-1} \bm{H}_1  \right\}\nonumber \\
&+{\rm{Diag}}\left\{ \bm{H}_2 \bm{H}_0^{\rm{H}} \left[\bm{\Sigma}^{-1}
\bm{H} \bm{H}^{\rm{H}}
+\bm{I}_{N_r}\right]^{-1} \bm{\Sigma}^{-1} \bm{H}_1  \right\} .
\end{align}
Unfortunately, even though the first-order derivative is derived, it is still difficult to derive the optimal closed-form solution from (\ref{diag_irs_log_deri}),
since its involved quadratic term w.r.t. $\bm{\Lambda}_{\bm{\Theta}}$ appears in an inverse form.
As a remedy, we intend to solve it based on problem transformation. 
Specifically, via introducing a series of auxiliary variables, \textbf{Prob.3} is equivalently transformed into
\begin{align}
\textbf{Prob.4:} \ \min_{{\bm{G}_{\bm{\Lambda}}}, \bm{\Lambda}_{\bm{\Theta}}, \bm{W}} 
\ &  {\rm{Tr}}\left(\bm{W}\left[ \bm{G}_{\bm{\Lambda}}\bm{H}-\bm{I}_{N_t}
\right]\left[ \bm{G}_{\bm{\Lambda}}\bm{H}-\bm{I}_{N_t}\right]^{\rm{H}} \right)\nonumber \\
&+{\rm{Tr}}(\bm{W} \bm{G}_{\bm{\Lambda}}\bm{\Sigma}\bm{G}_{\bm{\Lambda}}^{\rm{H}})-\log|\bm{W}| \nonumber \\
{\rm{s.t.}} \ & {\rm{Tr}}(\bm{\Lambda}_{\bm{\Theta}}\bm{\Lambda}_{\bm{\Theta}}^{\rm{H}}) \le K.
\end{align} 
The equivalence between \textbf{Prob.3} and \textbf{Prob.4} is built based on the idea of weighted MSE minimization (WMMSE) \cite{QShi_TSP_MMSE,Christensen2008}. 
Then, \textbf{Prob.4} can be efficiently solved via the AO among $\bm{G}_{\bm{\Lambda}}$, $\bm{\Lambda}_{\bm{\Theta}}$ and $\bm{W}$. 
Specifically, both optimal $\bm{G}_{\bm{\Lambda}}$ and $\bm{W}$ can be directly derived by taking the first-order derivatives of the objective function of \textbf{Prob.4} w.r.t. $\bm{G}_{\bm{\Lambda}}$ and $\bm{W}$ to zeros,
i.e.,
 $\bm{G}_{\bm{\Lambda}} = \bm{H}^{\rm{H}} \left(\bm{\Sigma}+\bm{H} \bm{H}^{\rm{H}} \right)^{-1},
 \bm{W}=\left(\bm{I}_{N_t}-\bm{G}_{\bm{\Lambda}}\bm{H} \right)^{-1}$ .
Then, the optimization problem w.r.t. $\bm{\Lambda}_{\bm{\Theta}}$ can be written as
\begin{align}
\textbf{Prob.5:} \ \min_{\bm{\Lambda}_{\bm{\Theta}}} \ 
&{\rm{Tr}}\left(\bm{W}\left[ \bm{G}_{\bm{\Lambda}}\bm{H}-\bm{I}_{N_t}
\right]\left[ \bm{G}_{\bm{\Lambda}}\bm{H}-\bm{I}_{N_t}\right]^{\rm{H}} \right) \nonumber \\
{\rm{s.t.}} \ & {\rm{Tr}}(\bm{\Lambda}_{\bm{\Theta}}\bm{\Lambda}_{\bm{\Theta}}^{\rm{H}}) \le K.
\end{align}
The first-order derivative of the objective function of \textbf{Prob.5} w.r.t. $\bm{\Lambda}_{\bm{\Theta}}$ is given by
\begin{align}
& {\rm{Diag}}\left\{\frac{\partial 
{\rm{Tr}}\left(\bm{W}\left[ \bm{G}_{\bm{\Lambda}}\bm{H}-\bm{I}_{N_t}
\right]\left[ \bm{G}_{\bm{\Lambda}}\bm{H}-\bm{I}_{N_t}\right]^{\rm{H}} \right) 
}{\partial \bm{\Lambda}_{\bm{\Theta}}}\right\} \nonumber \\
=& {\rm{Diag}}\{  \bm{H}_2\left(\bm{H}^{\rm{H}}\bm{G}_{\bm{\Lambda}}^{\rm{H}}
- {\bm{H}}_2\right)\bm{W}\bm{G}_{\bm{\Lambda}}\bm{H}_1\},
\end{align}
based on which the KKT optimality conditions of \textbf{Prob.5} can be formulated as
\begin{subequations}
  \begin{numcases}{}
   \!\! {\rm{Diag}}\{  \bm{H}_2\!\left(\!\bm{H}^{\rm{H}}\bm{G}_{\bm{\Lambda}}^{\rm{H}}
    \!-\! {\bm{H}}_2\!\right)\!\bm{W}\bm{G}_{\bm{\Lambda}}\bm{H}_1\}\!=\!\!-\mu {\rm{Diag}} \{ \bm{\Lambda}_{\bm{\Theta}}^{\rm{H}} \},\label{Prob5_Condition_a}\\
\!\!\mu \left({\rm{Tr}}(\bm{\Lambda}_{\bm{\Theta}}\bm{\Lambda}_{\bm{\Theta}}^{\rm{H}}) - K\right) =0,\!\label{Prob5_Condition_b}
  \end{numcases}
\end{subequations}
where $\mu$ is the dual variable associated with the amplitude constraint.
Based on (\ref{Prob5_Condition_a}), the optimal $\bm{\Lambda}_{\bm{\Theta}}$ is derived as
\begin{align}\label{opt_irs_diag_log}
{\rm{Diag}}\{\bm{\Lambda}_{\bm{\Theta}}^{\rm{H}}\}=(\mu{\bm{I}_K}+\bm{\Phi})^{-1}\bm{a},
\end{align} 
where $\bm{\Phi}\!\!\!=\!\!\bm{H}_2 \bm{H}_2^{\rm{H}} \!\odot\!
\left( \bm{H}_1^{\rm{H}}\bm{G}^{\rm{H}}\bm{W}\bm{G} \bm{H}_1\right)^{\rm{T}}$
and $\bm{a}\!\!=\!\!{\rm{Diag}}\{ \!{\bm{H}}_2\bm{W}\bm{G}_{\bm{\Lambda}}\bm{H}_1
\!\!-\!\!\bm{H}_2\bm{H}_0^{\rm{H}}\bm{G}_{\bm{\Lambda}}^{\rm{H}}
\bm{W}\bm{G}_{\bm{\Lambda}}\bm{H}_1\!\}$.
Moreover, since ${\rm{Tr}}(\bm{\Lambda}_{\bm{\Theta}}\bm{\Lambda}_{\bm{\Theta}}^{\rm{H}})$ is monotonically decreasing w.r.t. $\mu$, the optimal $\mu$ satisfying (\ref{Prob5_Condition_b}) is found via
the bisection search.

In addition, we formulate the MSE minimization problem for the amplitude-adjustable IRS-aided MIMO system as
\begin{align}
\textbf{Prob.6:} \ \min_{\bm{\Lambda}_{\bm{\Theta}}} \ & {\rm{Tr}}\left( \left[\bm{\Sigma}^{-1}\bm{H}\bm{H}^{\rm{H}}+\bm{I}_{N_r}\right]^{-1} \right) \nonumber \\
{\rm{s.t.}} \ & {\rm{Tr}}(\bm{\Lambda}_{\bm{\Theta}}\bm{\Lambda}_{\bm{\Theta}}^{\rm{H}}) \le K.
\end{align}
Similar to \textbf{Prob.3}, \textbf{Prob.6} is difficult to solve since it involves a quadratic term w.r.t. $\bm{\Lambda}_{\bm{\Theta}}$ appears in an inverse form.
Fortunately, it can also be equivalently transformed into the WMMSE minimization problem \textbf{Prob.4} by setting $\bm{W}=\bm{I}_{N_t}$.

\vspace{-2mm}
\subsection{Extension to Block-Diagonal Structure Constraints}
In this subsection, we extend the complex matrix derivative to the block-diagonal matrix, which is essentially a kind of bidiagonal matrix. 
In a general MU-MIMO uplink system, the received signal at the BS can be written as\cite{Serbetli2004}
\begin{align}\label{mu_channel}
\bm{y}=\bm{H}\bm{s}+\bm{n} \ \ \text{with} \ \
 \bm{H}=[\bm{H}_1,\cdots,\bm{H}_K],
\end{align}
where $\bm{H}_k\in \mathbb{C}^{N_t \times N_r}$ denotes the channel between the BS and the $k$th user, 
$\bm{s}\in \mathbb{C}^{N_r  \times 1}$ denotes the transmitted data stream to the $k$th user, 
and all data streams $\bm{s}_k$'s are stacked into the vector $\bm{s}\in \mathbb{C}^{N_r K \times 1}$, 
i.e., 
$\bm{s}=[ \bm{s}_1^{\rm{T}},\cdots,\bm{s}_K^{\rm{T}} ]^{\rm{T}}$.
Accordingly, the covariance matrix of $\bm{s}$ is a block-diagonal matrix, i.e.,
\begin{align}
\bm{Q}=\mathbb{E}\{\bm{s}{\bm{s}}^{\rm{H}}\}={\rm{Blockdiag}}\left(\left\{ \bm{Q}_k \right\}_{k=1}^K\right),
\end{align}
where $ \bm{Q}_k\!\!\!\!=\!\!\mathbb{E}\{\bm{s}_k{\bm{s}}_k^{\rm{H}}\}$ is the transmit covariance matrix of $\bm{s}_k$.
Hereafter, we mainly consider the capacity maximization problem under the power constraints as an example to introduce the application of our proposed complex matrix derivatives, 
which is formulated as
\begin{align}\label{MU-MIMO-Uplink}
\textbf{Prob.7:} \ \max_{\bm{Q}} \ & \log\left|\bm{I}_{N_t}+\bm{\Sigma}^{-1}\bm{H}\bm{Q}\bm{H}^{\rm{H}}\right| \nonumber \\
{\rm{s.t.}} \ & {\rm{Tr}}(\bm{Q}_k) \le P_k, \bm{Q}_k \succeq \bm{0},\ \forall k .
\end{align}
The differential of the objective function of \textbf{Prob.7} w.r.t. $\bm{Q}$ is given by
 ${\rm d}\left(\log\big|\bm{I}_{N_t}+\bm{\Sigma}^{-1}\bm{H}\bm{Q}\bm{H}^{\rm H}\big|\right) 
 = {\rm Tr}\big(
  \big(\bm{I}_{N_t}+\bm{\Sigma}^{-\frac{1}{2}}\bm{H}\bm{Q}\bm{H}^{\rm H}\bm{\Sigma}^{-\frac{1}{2} }\big)^{-1}
  \bm{\Sigma}^{-\frac{1}{2}}\bm{H}{\rm d}\big(\bm{Q}\big)\bm{H}^{\rm H}\bm{\Sigma}^{-\frac{1}{2}}\big)$.
The corresponding first-order derivative w.r.t $\bm{Q}$ is then derived as
\begin{align}
&\frac{\partial \log\big|\bm{I}_{N_t}+\bm{\Sigma}^{-1}\bm{H}\bm{Q}\bm{H}^{\rm H}\big|}{\partial \bm{Q}}  \nonumber\\ 
=&{\rm{Blockdiag}}\Big(\left\{ \bm{H}_k^{\rm{H}}\bm{\Sigma}^{-\frac{1}{2}}\big(\bm{I}_{N_t}+\bm{\Sigma}^{-\frac{1}{2}}\bm{H}\bm{Q}\bm{H}^{\rm H}\bm{\Sigma}^{-\frac{1}{2}}\big)^{-1}\right.\nonumber\\
&\left.\qquad\qquad \quad \times\bm{\Sigma}^{-\frac{1}{2}} \bm{H}_k\right\}_{k=1}^K\Big).
\end{align}
Following that, the KKT optimality conditions are given by
\begin{subequations}
  \begin{numcases}{}
\!{\rm{Blockdiag}}\bigg(\!\!\left\{ \bm{H}_k^{\rm{H}}\bm{\Sigma}^{-\frac{1}{2}}\big(\bm{I}_{N_t}\!+\!\bm{\Sigma}^{-\frac{1}{2}}\bm{H}\bm{Q}\bm{H}^{\rm H}\bm{\Sigma}^{-\frac{1}{2}}\big)^{-1}\right.\nonumber\\
\left.\ \times
\bm{\Sigma}^{-\frac{1}{2}}\!\bm{H}_k \right\}_{k=1}^K\!\! \bigg)
\!=\!{\rm{Blockdiag}}\!\left(\!\left\{ \mu_k\bm{I}_{N_r}\!-\!\bm{\Psi}_k \right\}_{k=1}^K\!\right)\!,\!  \label{Prob_14_Condition_a}\\
\mu_k \left( {\rm{Tr}}(\bm{Q}_k) - P_k\right)=0,\label{Prob_14_Condition_b}\\
 {\rm{Tr}}(\bm{\Psi}_k \bm{Q}_k) = 0,\ \forall k,
  \end{numcases}
\end{subequations}
where $\mu_k$ and $\bm{\Psi}_k$ are the Lagrange multipliers corresponding to the transmit power constraint and the positive semi-definite constraint at the $k$th user, respectively.
Then, (\ref{Prob_14_Condition_a}) can be rewritten in terms of $\bm{Q}_k$ as follows:
\begin{align}
&\bm{H}_k^{\rm{H}}\bm{\Sigma}^{-\frac{1}{2}}\big(\bm{I}_{N_t}+\bm{\Sigma}^{-\frac{1}{2}}\bm{H}\bm{Q}\bm{H}^{\rm H}\bm{\Sigma}^{-\frac{1}{2}}\big)^{-1}\bm{\Sigma}^{-\frac{1}{2}}\bm{H}_k\nonumber\\
 = &\bm{H}_k^{\rm{H}}\bm{\Sigma}^{-\frac{1}{2}}\bm{L}_k^{-\frac{1}{2}}
 \big(\bm{I}_{N_t} +  \bm{L}_k^{-\frac{1}{2}}\bm{\Sigma}^{-\frac{1}{2}}\bm{H}_k\bm{Q}_k\bm{H}_k^{\rm H}\bm{\Sigma}^{-\frac{1}{2}}\bm{L}_k^{-\frac{1}{2}}\big)^{-1}\nonumber\\
 & \times\bm{L}_k^{-\frac{1}{2}}\bm{\Sigma}^{-\frac{1}{2}}\bm{H}_k
 =\mu_k\bm{I}_{N_r}-\bm{\Psi}_k, \ \forall k,
\end{align} 
where $\bm{L}_k = \bm{I}_{N_t}
 +{\sum}_{j\not{=}k}\bm{\Sigma}^{-\frac{1}{2}}\bm{H}_j\bm{Q}_j\bm{H}_j^{\rm H}\bm{\Sigma}^{-\frac{1}{2}}$.
Thus, based on eigenspace alignment, the optimal $\bm{Q}_k$'s can be derived as that in [16, Theorem 1],
i.e.,
$\bm{Q}_{k}=\bm{V}_{\bm{\mathcal{H}}_k}{\bm{\Lambda}}_{{\bm{Q}}_k}
{\bm{V}}_{\bm{\mathcal{H}}_k}^{\rm{H}}, \ \forall k$,
where ${\bm{\Lambda}}_{{\bm{Q}}_k}$ is a diagonal matrix, each diagonal element of which has a water-filling form,
and $\bm{V}_{\bm{\mathcal{H}}_k}$ is an unitary matrix coming from the singular value decomposition (SVD) represented as
$ \bm{L}_k^{-\frac{1}{2}}\bm{\Sigma}^{-\frac{1}{2}}\bm{H}_k
=\bm{U}_{\bm{\mathcal{H}}_k}{\bm{\Lambda}}_{\bm{\mathcal{H}}_k}
{\bm{V}}_{\bm{\mathcal{H}}_k}^{\rm{H}}  \ \text{with} \ {\bm{\Lambda}}_{\bm{\mathcal{H}}_k} \searrow$,
where $\bm{\Lambda}_{\bm{\mathcal{H}}_k} \searrow$ implies that the diagonal elements of $\bm{\Lambda}_{\bm{\mathcal{H}}_k}$ are arranged in descending order.
Similarly, $\mu_k$ that satisfying (\ref{Prob_14_Condition_b}) can be obtained by the bisection search.

\noindent\textit{Remark 1:} Based on the above discussions, we can conclude  that  globally optimal solutions of several classical optimization problems in the state-of-the-art wireless systems can be directly obtained with low complexity  by  the proposed complex matrix derivatives under  diagonal structure constraints.
In addition, for optimization problems that not satisfy diagonal structure constraints directly, the proposed algorithm is also able to obtain an approximate solution by further exploring the inherent structure of the optimal solution. 
For  example,
 the optimal matrix variables in the point-to-point MIMO system operating at high SNR conditions 
 and the MU-MISO downlink system employing the BD-ZF strategy \cite{zf_diag} are both valiated to be approximately diagonal.

\vspace{-2mm}
\section{Constant Modulus Constraints}\label{sec_modulus}

\begin{table*}[t]
  \renewcommand\arraystretch{1.25}
  \vspace{-12mm}
	\centering
	\caption{Element-Wise Phase Derivatives under The Constant Modulus Constraint}  \label{tab2}
	\vspace{-2mm}
  \resizebox{\textwidth}{!}{
		\begin{tabular}{|p{5.6cm}|p{9.6cm}|}
			\hline
			\textbf{Function Type}  & \textbf{Element-Wise Phase Derivative w.r.t. }$[\bm{\Theta}]_{i,j}, \forall i, j$ \\
			\hline
      $f_{\rm{C,TL}}={\rm{Tr}}({\bm{B}}^{\rm{H}}\bm{X})+{\rm{Tr}}({\bm{B}}\bm{X}^{\rm{H}})$
      & $-2\Im\left\{ [\bm{B}^*]_{i,j}[\bm{X}]_{i,j}\right\}$\\
			\hline
      $f_{\rm{C,TQ}}={\rm{Tr}}(\bm{X}{\bm{\Pi}}\bm{X}^{\rm{H}}\bm{\Phi}) $
      & $-2\Im\left\{ [(\bm{\Phi}\bm{X}\bm{\Pi})^*]_{i,j}[\bm{X}]_{i,j}\right\}$  \\
      \hline
      $f_{\rm{C,TI}}= {\rm Tr} \left((\bm{\Phi}+\bm{X}^{\rm{H}}\bm{\Pi}\bm{X})^{-1}\right)$
      &$2\Im\left\{ \left[ \left(\bm{\Pi}
      \bm{X}(\bm{\Phi}+\bm{X}^{\rm{H}}\bm{\Pi}\bm{X})^{-2}
      \right)^*\right]_{i,j} [\bm{X}]_{i,j}\right\}$\\
      \hline
      $f_{\rm{C,LD}}=\log|\bm{\Phi}+\bm{X}^{\rm{H}}\bm{\Pi}\bm{X}|$
      &$-2\Im\left\{ \left[ \left(\bm{\Pi}
      \bm{X}(\bm{\Phi}+\bm{X}^{\rm{H}}\bm{\Pi}\bm{X})^{-1}
      \right)^*\right]_{i,j} [\bm{X}]_{i,j}\right\}$\\
      \hline
	\end{tabular}}
  \vspace{-5mm}
\end{table*}

Different from diagonal structure constraints, 
constant modulus constraints are imposed on matrix variables in an element-wise manner, 
which makes the  optimization problem challenging to directly solve using complex matrix derivatives.
Motivated by this fact, 
we firstly provide some mathematical preliminaries for the element-wise phase derivatives of several widely adopted objective functions. Then, we investigate  specific optimization problems in both the hybrid analog-digital MIMO system and the fully-passive IRS-aided MIMO system.
In  order to avoid complicated matrix inversion and matrix factorization, a novel  AO algorithm with the aid of several arbitrary feasible solutions is proposed.
\vspace{-3.5mm}
\subsection{Mathematical Preliminaries}\label{unit_modu_math}
We firstly introduce a complex matrix variable $\bm{X} \in \mathbb{C}^{N \times M}$ subject to constant modulus constraints as follows:
\begin{align}\label{def_cons}
[\bm{X}]_{i,j} =e^{\jmath\theta_{i,j}} , \ \forall i, j\ \text{and} \ {\rm{Tr}}(\bm{X}\bm{X}^{\rm{H}})=NM,
\end{align}
where $\theta_{i,j} \in [0,2\pi]$ denotes the phase of $[\bm{X}]_{i,j}$.
Based on (\ref{def_cons}), the first-order derivative w.r.t. the constant modulus constrained $\bm{X}$ can be replaced by the first-order derivative w.r.t. the corresponding unconstrained phase matrix $\bm{\Theta}$, 
where $[\bm{\Theta}]_{i,j}=\theta_{i,j}, \forall i,j$. 
Accordingly, the element-wise phase derivatives of the function $f(\bm{X})$ w.r.t. $\bm{\Theta}$ can be defined as
\begin{align}
\left[\frac{\partial f(\bm{X})}{\partial\bm{\Theta}}\right]_{i,j}
=\frac{\partial f(\bm{X})}{\partial [\bm{\Theta}]_{i,j}},
 \ \forall i, j.
\end{align} 
Similar to Sec. \ref{sec_diag_math}, we also consider the element-wise phase derivatives for four common objective functions, 
i.e., 

\subsubsection{ Trace-Linear Function}

Since the phase $[\bm{\Theta}]_{i,j}$'s are real scalar,
for arbitrary complex matrix $\bm{B} \in \mathbb{C}^{N \times M}$, the element-wise phase derivatives of a trace-linear function
$f_{\rm{C,TL}}={\rm{Tr}}({\bm{B}}^{\rm{H}}\bm{X})+{\rm{Tr}}({\bm{B}}\bm{X}^{\rm{H}})$
w.r.t. $[\bm{\Theta}]_{i,j}$'s can be obtained as
\begin{align}
\left[\frac{\partial f_{\rm{C,TL}}}{\partial \bm{\Theta}}\right]_{i,j}
=&\left[\frac{\partial \sum_m\sum_n[\bm{B}^*]_{n,m}[\bm{X}]_{n,m}}{\partial \bm{\Theta}}\right]_{i,j}\nonumber\\
&+\left[\partial \frac{\sum_m\sum_n[\bm{B}]_{n,m}[\bm{X}^*]_{n,m}}{\partial \bm{\Theta}}\right]_{i,j}\nonumber\\
=&\jmath[\bm{B}^*]_{i,j}[\bm{X}]_{i,j} - \jmath[\bm{B}]_{i,j}[\bm{X}^*]_{i,j}\nonumber\\
=&-2\Im\!\left\{ [\bm{B}^*]_{i,j}[\bm{X}]_{i,j}\!\right\}\!, \ \forall i, j.
\end{align}

\subsubsection{Trace-Quadratic Function}
For arbitrary Hermitian matrices $\bm{\Pi}\in \mathbb{C}^{M \times M}$ and $\bm{\Phi}\in \mathbb{C}^{N \times N}$, the element-wise phase derivatives of a trace-quadratic function $f_{\rm{C,TQ}}={\rm{Tr}}(\bm{X}{\bm{\Pi}}\bm{X}^{\rm{H}}\bm{\Phi})$
w.r.t. $[\bm{\Theta}]_{i,j}$'s are given by
\begin{align}
\left[\frac{\partial f_{\rm{C,TQ}} }{\partial \bm{\Theta}}\right]_{i,j}&=\jmath[(\bm{\Phi}\bm{X}\bm{\Pi})^*]_{i,j}[\bm{X}]_{i,j}-
\jmath[(\bm{\Phi}\bm{X}\bm{\Pi})]_{i,j}[\bm{X}^*]_{i,j} \nonumber \\
&=-2\Im\left\{ [(\bm{\Phi}\bm{X}\bm{\Pi})^*]_{i,j}[\bm{X}]_{i,j}\right\}, \ \forall i, j.
\end{align} 

\subsubsection{Trace-Inverse Function}
Regarding a trace-inverse function $f_{\rm{C,TI}}={\rm{Tr}}\left((\bm{\Phi}+\bm{X}^{\rm{H}}\bm{\Pi}\bm{X})^{-1}\right)$, 
we have the following element-wise phase derivatives w.r.t. $[\bm{\Theta}]_{i,j}$'s.
\begin{align}
  \frac{\partial f_{\rm{C,TI}} }
  {\partial\bm{\Theta}}
  =&-\jmath\left[ \left( \bm{\Phi}+\bm{X}^{\rm{H}}\bm{\Pi}\bm{X} \right)^{-2}\bm{X}^{\rm{H}}\bm{\Pi}\right]_{i,j} [\bm{X}]_{i,j}\nonumber\\
  &+\jmath\left[ \left(\bm{\Pi}
   \bm{X}(\bm{\Phi}+\bm{X}^{\rm{H}}\bm{\Pi}\bm{X})^{-2}
   \right)\right]_{i,j} [\bm{X}^*]_{i,j}  \\
    =& 2\Im\left\{ \left[ \left(\bm{\Pi}
   \bm{X}(\bm{\Phi}+\bm{X}^{\rm{H}}\bm{\Pi}\bm{X})^{-2}
   \right)^*\right]_{i,j}\! [\bm{X}]_{i,j}\right\}, \ \forall i, j.\nonumber
  \end{align}

\subsubsection{Log-Determinant Function}
Similarly, the element-wise phase derivatives of a log-determinant function $f_{\rm{C,LD}}=\log|\bm{\Phi}+\bm{X}^{\rm{H}}\bm{\Pi}\bm{X}|$
w.r.t. $[\bm{\Theta}]_{i,j}$'s are given by
\begin{align}\label{deri_log_def}
  \left[\frac{\partial f_{\rm{C,LD}}}{\partial \bm{\Theta}}\right]_{i,j}
   =& -2\Im\biggl\{  \left[ \left(\bm{\Pi}
   \bm{X}(\bm{\Phi}+\bm{X}^{\rm{H}}\bm{\Pi}\bm{X})^{-1}
   \right)^*\right]_{i,j}\nonumber\\
  &\qquad \quad \times [\bm{X}]_{i,j}\biggr\}, \ \forall i, j.
  \end{align}

The element-wise phase derivatives of the above four types of objective functions are summarized in Table \ref{tab2}.
In the following subsection, several state-of-the-art wireless applications will be investigated in detail based on the above fundamental properties.

\begin{figure*}[t]
  \vspace{-12mm}
    \centering
    \includegraphics[width=0.9\textwidth]{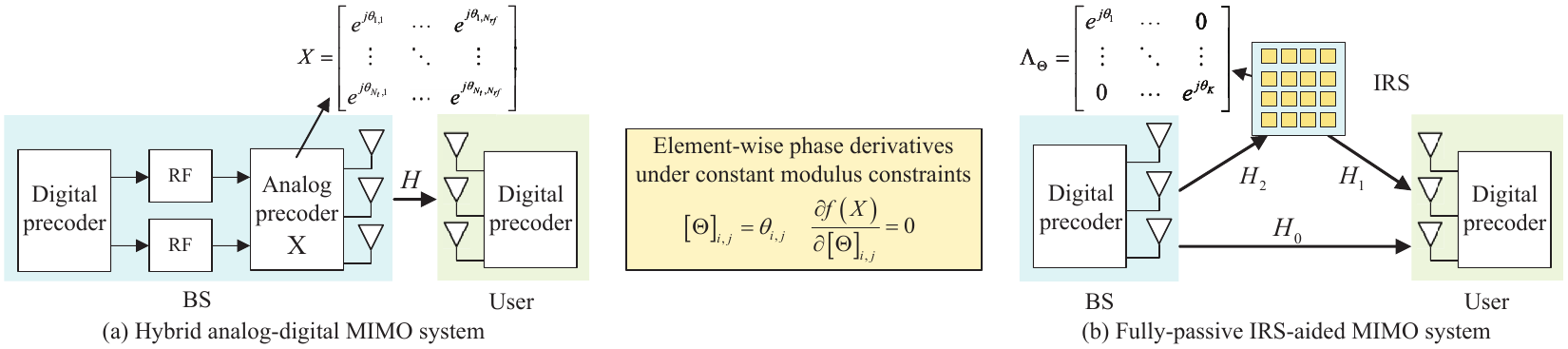}
	\vspace{-3mm}
    \caption{A  diagram of  application  scenarios associated  with  constant modulus matrix variables.}
   \label{fig_sys_cons}
    \vspace{-5mm}
\end{figure*}

\vspace{-3mm}
\subsection{Specific Wireless Applications}\label{sec_const_app}
In Fig. \ref{fig_sys_cons}, there are two typical wireless applications associated with constant modulus constraints, 
i.e., the analog beamforming optimization in the hybrid analog-digital MIMO system and the phase shift optimization in the fully-passive IRS-aided MIMO system,
which are elaborated as follows.

\subsubsection{Hybrid Analog-Digital MIMO System}
We firstly consider the hybrid analog-digital beamforming optimization design in the downlink point-to-point MIMO system
\cite{XYu_JSTSP_Hybrid}. 
Then, the received signal at the user can be expressed as
\begin{align}\label{sys_cons_hy}
 \bm{y}=\bm{G}\bm{H}\bm{X} \bm{F}_{\rm{D}}\bm{s}
 +\bm{G}\bm{n} ,
\end{align}
where $\bm{G}\!\in \!\mathbb{C}^{N_{s} \times N_{r}}$ denotes the fully-digital receive equalizer,
$\bm{H}\!\in \!\mathbb{C}^{N_{r} \times N_{t}}$ denotes the channel between the BS and the user,
$\bm{X}\!\in\! \mathbb{C}^{N_t \times N_{rf}}$ and $\bm{F}_{\rm{D}}\!\in\! \mathbb{C}^{N_{rf} \times N_{s}}$ are the constant modulus analog beamformer and the digital beamformer, respectively.
$\bm{s}\!\in \!\mathbb{C}^{N_{s} \times 1}$ is the transmit data streams with unit covariance matrix, i.e., $\mathbb{E}\{\bm{s}{\bm{s}}^{\rm{H}}\}\!=\!\bm{I}_{N_s}$.
$\bm{n}\!\in\! \mathbb{C}^{N_{r} \times 1}$ is the additive Gaussian noise with zero mean and covariance matrix $\mathbb{E}\{\bm{n}\bm{n}^{\rm{H}}\}\!=\!\bm{\Sigma}$.
Based on (\ref{sys_cons_hy}), the MSE matrix  is given by
\begin{align}
  \bm{E}_{\rm{MSE}}=&\mathbb{E}\left[\left(\hat{\bm{s}}-\bm{s}\right)\left(\hat{\bm{s}}-\bm{s}\right)^{\rm{H}}\right]\nonumber\\ 
  =&\left(\bm{G} \bm{H} \bm{X} \bm{F}_{\rm{D}}-\bm{I}_{N_s} \right) \left(\bm{G} \bm{H} \bm{X} \bm{F}_{\rm{D}}-\bm{I}_{N_s}\right)^{\rm{H}} 
  + \bm{G}\bm{\Sigma}\bm{G}^{\rm{H}} \nonumber\\
  \overset{(c)}{=}& \left[\bm{I}_{N_s}+   \bm{F}_{\rm{D}}^{\rm{H}} \bm{X}^{\rm{H}}\bm{H}^{\rm{H}} \bm{\Sigma}^{-1} \bm{H} \bm{X} \bm{F}_{\rm{D}}\right]^{-1}, 
\end{align} 
where $\hat{\bm{s}}$ denotes the estimated signal 
and $(c)$ holds based on the optimal unconstrained Wiener filter $\bm{G}\!=\!\bm{F}_{\rm{D}}^{\rm{H}} \bm{X}^{\rm{H}} \bm{H}^{\rm{H}}\left(\bm{H} \bm{X} \bm{F}_{\rm{D}} \bm{F}_{\rm{D}}^{\rm{H}} \bm{X}^{\rm{H}} \bm{H}^{\rm{H}}\!+\!\bm{R}_{\rm{n}}\right)^{-1}$ \cite{Palomar_weiner}.
Without loss of generality, we usually assume $\bm{F}_{\rm{D}}\bm{F}_{\rm{D}}^{\rm{H}}\approx \gamma^2 \bm{I}_{N_{rf}}$ for large-scale MIMO systems\cite{Wei_Yu_Hybrid}.
Under this assumption, the capacity maximization problem of the hybrid analog-digital MIMO system can be formulated as
\begin{align}
\textbf{Prob.8:} \ \max_{\bm{X}} \ & \log|\bm{I}_{N_{rf}}+\bm{X}^{\rm{H}}\bm{\Pi}\bm{X}| \nonumber \\
{\rm{s.t.}}  \ & |[\bm{X}]_{i,j}|=1, \ \forall i, j,
\end{align} 
where $\bm{\Pi}=\gamma^2 \bm{H}^H \bm{\Sigma}^{-1} \bm{H}$ represents the effective signal-to-noise ratio (SNR).
According to the KKT optimality conditions, the element-wise phase derivatives of the objective function of \textbf{Prob.8} w.r.t. $[\bm{X}]_{i,j}$'s must equal zeros at the optimal $[\bm{X}]_{i,j}$'s.
Specifically, we recall (\ref{deri_log_def}) to obtain 
\begin{align}\label{g_12}
   \underbrace{\Im\!\left\{ \left[ \left(\bm{\Pi}
 \bm{X}(\bm{I}_{N_{rf}}\!+\!\bm{X}^{\rm{H}}\bm{\Pi}\bm{X})^{-1}
 \right)^*\right]_{i,j}\! \![\bm{X}]_{i,j}\right\}}_{\triangleq g_{i,j}(\bm{X})}\!=\!0, \ \forall i, j.
\end{align} 
Define $\bm{A}_j\!=\!\bm{I}_{N_t}+\bm{\widetilde X}_j\bm{\widetilde X}_j^{\rm{H}}\bm{\Pi}$  
and $ n_j=1+[\bm{X}]_{:,j}^{\rm{H}}\bm{\Pi}\bm{A}_j^{-1}[\bm{X}]_{:,j}$,
where $\bm{\widetilde X}_{j}$ denotes the sub-matrix of $\bm{X}$ with the $j$th column removed.
Then, $\left[\bm{\Pi}
 \bm{X}(\bm{I}_{N_{rf}}+\bm{X}^{\rm{H}}\bm{\Pi}\bm{X})^{-1}
 \right]_{i,j}$'s can be rewritten as
\begin{align}\label{factor_x_hyb_log}
& \left[ \bm{\Pi}
 \bm{X}(\bm{I}_{N_{rf}}+\bm{X}^{\rm{H}}\bm{\Pi}\bm{X})^{-1}
 \right]_{i,j} \\
 \overset{(d)}{=} &\left[\bm{\Pi}\left(\bm{A}_j^{-1}-\frac{\bm{A}_j^{-1}[\bm{X}]_{:,j}[\bm{X}]_{:,j}^{\rm{H}}
 \bm{\Pi}\bm{A}_j^{-1}}{n_j}\right)
 \bm{X} \right]_{i,j}\nonumber\\
 = &\frac{1}{n_j}{\sum}_{l\not{=}i}
 [\bm{\Pi}\bm{A}_j^{-1}]_{l,i}^{\rm H}[\bm{X}]_{l,j}
 +\frac{1}{n_j}[\bm{\Pi}\bm{A}_j^{-1}]_{i,i}^{\rm H}[\bm{X}]_{i,j}, \ \forall i, j,\nonumber
\end{align} 
where $(d)$ holds similarly to (\ref{left_diag_log}).
By substituting (\ref{factor_x_hyb_log}) into (\ref{g_12}), we have
\begin{subequations}
\begin{align}\label{gen_hy_log}
  &\Im\left\{ \left[ \left(\bm{\Pi}
 \bm{X}(\bm{I}_{N_{rf}}+\bm{X}^{\rm{H}}\bm{\Pi}\bm{X})^{-1}
 \right)^*\right]_{i,j} [\bm{X}]_{i,j}\right\}\\
 =&\Im \left\{ \left(\frac{1}{n_j}{\sum}_{l\not{=}i}
 [\bm{\Pi}\bm{A}_j^{-1}]_{l,i}^{\rm H}[\bm{X}]_{l,j}\right)^*  [\bm{X}]_{i,j}\right.\nonumber\\
 &\left.\quad+\left(\frac{1}{n_j}[\bm{\Pi}\bm{A}_j^{-1}]_{i,i}^{\rm H}\right)^* \right\}=0,\ \forall i, j.\label{gen_hy_log_b} 
\end{align}
\end{subequations}
Since the following equality holds, i.e.,
\begin{align}\label{PiA_her}
 \bm{\Pi}\bm{A}_j^{-1}=&\bm{\Pi}(\bm{I}_{N_t}+\bm{\widetilde X}_j\bm{\widetilde X}_j^{\rm{H}}\bm{\Pi})^{-1}\nonumber\\
 =&(\bm{I}_{N_t}+\bm{\Pi}\bm{\widetilde X}_j\bm{\widetilde X}_j^{\rm{H}})^{-1}\bm{\Pi}=(\bm{\Pi}\bm{A}_j^{-1})^{\rm{H}}, \ \forall  j,
\end{align}
we can conclude that $\bm{\Pi}\bm{A}_j^{-1}$ is a Hermitian matrix.
As such, it is readily inferred that  $n_j$ and $\frac{1}{n_j}[\bm{\Pi}\bm{A}_j^{-1}]_{i,i}^{\rm H}$ are both real scalars.
Recall the definition of $[\bm{X}]_{i,j}=e^{\jmath\theta_{i,j}},\forall i,j$, 
it follows from (\ref{gen_hy_log_b}) that the optimal $[\bm{\Theta}]_{i,j}$'s to \textbf{Prob.8} are derived as
\begin{align}
[\bm{\Theta}]_{i,j}
=&{\rm{Phase}}\left\{{\sum}_{l\not{=}i}
[\bm{\Pi}\bm{A}_j^{-1}]_{l,i}^{\rm H}[\bm{X}]_{l,j}
\right\} \nonumber\\ \text{or} \
&\pi + {\rm{Phase}}\left\{{\sum}_{l\not{=}i}
[\bm{\Pi}\bm{A}_j^{-1}]_{l,i}^{\rm H}[\bm{X}]_{l,j}
\right\}
, \ \forall i, j.
\end{align}

In addition, the MSE minimization problem is expressed  as
\begin{align}
\textbf{Prob.9:} \ \min_{\bm{X}} \ & {\rm Tr} \left(\left(\bm{I}_{N_{rf}}+\bm{X}^{\rm{H}}\bm{\Pi}\bm{X}\right)^{-1}\right) \nonumber \\
{\rm{s.t.}} \ & |[\bm{X}]_{i,j}|=1,\ \forall i, j.
\end{align}
Similarly, to find the optimal solution of \textbf{Prob.9}, 
the element-wise phase derivatives of the objective function w.r.t. $[\bm{X}]_{i,j}$'s must equal zeros, that is,
\begin{align}\label{g_13}
 \underbrace{ \Im\!\left\{\! \left[ \left(\bm{\Pi}
 \bm{X}\left(\!\bm{I}_{N_{rf}}\!+\!\bm{X}^{\rm{H}}\bm{\Pi}\bm{X}\!\right)^{-2}
 \right)^*\right]_{i,j}\!\! [\bm{X}]_{i,j}\right\}}_{\triangleq g_{i,j}(\bm{X})}\!=\!0, \ \forall i, j.
\end{align} 
The terms $\left[ \bm{\Pi}
\bm{X}\!\left(\bm{I}_{N_{rf}}\!+\!\bm{X}^{\rm{H}}\bm{\Pi}\bm{X}\right)^{-2}
\right]_{i,j}$'s can be rewritten as
\begin{align}\label{factor_x_hyb_tr}
  & \left[\bm{\Pi}\bm{X}(\bm{I}_{N_{rf}}+\bm{X}^{\rm{H}}\bm{\Pi}\bm{X})^{-2}\right]_{i,j} \nonumber\\
  \overset{(e_1)}{=} &\left[\bm{\Pi}\left(\bm{A}_j^{-1}-\frac{\bm{A}_j^{-1}[\bm{X}]_{:,j}[\bm{X}]_{:,j}^{\rm{H}}
   \bm{\Pi}\bm{A}_j^{-1}}{n_j}\right)^2
   \bm{X} \right]_{i,j}\nonumber\\
  \overset{(e_2)}{=}&  \frac{1}{n_j}[\bm{\Pi}\bm{A}_j^{-2}]_{:,i}^{\rm H}[\bm{X}]_{:,j}-\frac{m_j}{n_j^2}[\bm{\Pi}\bm{A}_j^{-1}]_{:,i}^{\rm H}[\bm{X}]_{:,j}\nonumber\\
  \overset{(e_3)}{=}& \frac{A_{1,i,j}^*  + B_{1,i,j}^* [\bm{X}]_{i,j}^2 + C_{1,i,j}^* [\bm{X}]_{i,j}}{n_j^2}, \ \forall i, j,
  \end{align} 
where $(e_1)$ holds due to the same reasons as (\ref{left_diag_log}),
$(e_2)$ holds by defining $m_j=[\bm{X}]_{:,j}^{\rm{H}}\bm{\Pi}\bm{A}_j^{-2}[\bm{X}]_{:,j}$, which is a real scalar and this can be proved similarly to (\ref{PiA_her}).
$(e_3)$ is obtained by rewriting $n_j$ and $m_j$ in terms of $[\bm{X}]_{i,j}$ as
\begin{align}\label{njmj}
  &n_j \!=\! \zeta_j^n \!+\! 2 \Re \left\{ \eta_j^n [\bm{X}]_{i,j}^{*} \right\} , 
  m_j\! = \!\zeta_j^m \!+\! 2 \Re \left\{ \eta_j^m [\bm{X}]_{i,j}^{*} \right\} ,
\end{align}
where $
  \zeta_j^n\!=\!  1\!+\! [\bm{\Pi}\bm{A}_j^{-1}]_{i,i} 
  \!+\! \Re\!\left\{\!\sum_{p \neq i, q \neq i} [\bm{X}]_{p,j}^* 
   [\bm{\Pi}\bm{A}_j^{-1}]_{p,q}\notag \right.
   \\
   \left. [\bm{X}]_{q,j} \right\}$,
 $ \zeta_j^m\!=\!  [\bm{\Pi}\bm{A}_j^{-2}]_{i,i}
   \!+\! \Re\left\{\!\sum_{p \neq i, q \neq i} \![\bm{X}]_{p,j}^*
   [\bm{\Pi}\bm{A}_j^{-2}]_{p,q}\notag \right.
   \\
   \left. [\bm{X}]_{q,j} \right\}$,  
 $ \eta_j^n\!=  \!\sum_{l\not{=}i} [\bm{\Pi}\bm{A}_j^{-1}]_{l,i}[\bm{X}]_{l,j}^{*}$,
 $\eta_j^m\!=\! \sum_{l\not{=}i} [\bm{\Pi}\bm{A}_j^{-2}]_{l,i}  [\bm{X}]_{l,j}^{*}$ , $ \forall i, j$.
Moreover, $A_{1,i,j}$, $B_{1,i,j}$ and $C_{1,i,j}$ in (\ref{factor_x_hyb_tr}) are defined as
\begin{align}
  A_{1,i,j}=&(\eta_j^n)^*[\bm{\Pi}\bm{A}_j^{-2}]^*_{i,i} - (\eta_j^m)^*[\bm{\Pi}\bm{A}_j^{-1}]^*_{i,i} \nonumber\\
  &+ (\zeta_j^n)^*(\eta_j^m)^* - (\zeta_j^m)^*(\eta_j^n)^*,\nonumber\\
  B_{1,i,j}=&\eta_j^n[\bm{\Pi}\bm{A}_j^{-2}]^*_{i,i}-\eta_j^m [\bm{\Pi}\bm{A}_j^{-1}]^*_{i,i},\nonumber\\
  C_{1,i,j}=&(\zeta_j^n)^*[\bm{\Pi}\bm{A}_j^{-2}]^*_{i,i} - (\zeta_j^m)^*[\bm{\Pi}\bm{A}_j^{-1}]^*_{i,i}\nonumber\\
  & + \eta_j^n (\eta_j^m)^* - \eta_j^m (\eta_j^n)^* , \ \forall i, j.
\end{align}
By substituting (\ref{factor_x_hyb_tr}) into (\ref{g_13}), we have
\begin{subequations}
\begin{align}\label{gen_hy_tr}
& \Im \left\{  \left[ \left(\bm{\Pi}
 \bm{X}(\bm{I}_{N_{rf}}+\bm{X}^{\rm{H}}\bm{\Pi}\bm{X})^{-2}
 \right)^*\right]_{i,j} [\bm{X}]_{i,j} \right\} \\
 =& \Im \left\{\frac{A_{1,i,j} [\bm{X}]_{i,j} \!+\! B_{1,i,j} [\bm{X}]_{i,j}^* \!+\! C_{1,i,j}}{n_j^2} \right\}
 \!=\!0, \ \forall i, j.\label{gen_hy_tr_c}
\end{align}
\end{subequations}
Then, since $\Im\{ \frac{a}{b} \} =0$ is equal to $\Im\{ a \} =0$ for a real scalar $b$, (\ref{gen_hy_tr_c}) can be further simplified as
\begin{subequations}
\begin{align}\label{gen_hy_tr_d}
    &\Im \left\{ A_{1,i,j} e^{\jmath \theta_{i,j}} + B_{1,i,j} e^{-\jmath \theta_{i,j}} + C_{1,i,j} \right\} \\
    \overset{(f_1)}{=}&\!\!| A_{1,i,j}| \sin (\theta_{i,j}\!+\!\alpha_{i,j} ) \!-\! | B_{1,i,j}| \sin (\theta_{i,j}\! -\!\beta_{i,j}  )\! +\!\Im \{C_{1,i,j} \} \nonumber\\
    \overset{(f_2)}{=}&\!\!\!\sqrt{z_{1,i,j}^2\!+\!z_{2,i,j}^2} \sin (\theta_{i,j}\! +\!\phi_{i,j} )  \!+\!\Im \{C_{1,i,j} \}\!=\!0,\! \ \forall i, j,\label{gen_hy_tr_db}
  \end{align}
\end{subequations}
where $(f_1)$ is obtained using Euler's formula with $A_{1,i,j}\!=\!| A_{1,i,j}| e^{j \alpha_{i,j}}$ and $B_{1,i,j}\!=\!| B_{1,i,j}| e^{j \beta_{i,j}}$,
$(f_2)$ holds due to the  sum-to-product trigonometric identity with 
$z_{1,i,j}\!=\!| A_{1,i,j}|\cos(\alpha_{i,j})\!+\!| B_{1,i,j}| \cos (\beta_{i,j})$, $z_{2,i,j}\!=\!| A_{1,i,j}|\sin(\alpha_{i,j})\!+\!| B_{1,i,j}| \sin (\beta_{i,j})$ and
\begin{align}
	\phi_{i,j}=
    \begin{cases}
		\arctan \left(\frac{z_{2,i,j}}{z_{1,i,j}}\right), & \text { if } z_{1,i,j} \geq 0 , \\
		\pi-\arctan \left(\frac{z_{2,i,j}}{z_{1,i,j}}\right), & \text { if } z_{1,i,j}<0, 
		\end{cases}\ \forall i, j.
\end{align}
Based on (\ref{gen_hy_tr_db}), the optimal $\left[\bm{\Theta}\right]_{i,j}$'s to \textbf{Prob.9} are obtained as
\begin{align}\label{theta_log}
	\left[\bm{\Theta}\right]_{i,j}=& -\phi_{i,j}+\arcsin \left(-\frac{\Im \{C_{1,i,j} \}}{\sqrt{z_{1,i,j}^2+z_{2,i,j}^2}}\right) \\
  & \text{or}  \
  \pi-\phi_{i,j} +\arcsin\left(-\frac{\Im \{C_{1,i,j} \}}{\sqrt{z_{1,i,j}^2+z_{2,i,j}^2}}\right), \ \forall i, j.\nonumber
\end{align}

Next, we consider a general WMMSE minimization problem often studied in the hybrid MU-MIMO system, which can be formulated as\cite{SGong_mm}
\begin{align}
\textbf{Prob.10:} \ \min_{ \{\bm{X}_u\} } \ &  
\sum_{u=1}^{U}\left( 
{\rm{Tr}}(\bm{\Phi}_u\bm{X}_u\bm{\Pi}_u\bm{X}_u^{\rm{H}})-{\rm{Tr}}(\bm{B}_u^{\rm{H}}\bm{X}_u)\right.\nonumber\\
&\left.\qquad-{\rm{Tr}}(\bm{B}_u\bm{X}_u^{\rm{H}}) \right) \nonumber \\
{\rm{s.t.}}  \ & |[\bm{X}_u]_{i,j}|=1,\ \forall u,i, j,
\end{align} 
where $\bm{X}_u$'s are the analog beamformer for the $u$th user.
$\bm{\Phi}_u \!\!\!\in\!\!\! \mathbb{C} ^{N_t \times N_t}$'s, $\bm{\Pi}_u\!\!\!\in\! \!\mathbb{C} ^{N_{rf} \times N_{rf}}$'s and $\bm{B}_u\!\!\in\!\!\! \mathbb{C} ^{N_t \times N_{rf}}$'s denote the corresponding effective channel covariance matrix, the digital beamforming covariance matrix and the cascade channel, respectively,
which are mathematically modeled as
\begin{align}
  &\bm{\Phi}_u=\!
	\begin{cases}
		\bm{H}_{u,u}^{\rm{H}}\bm{G}_{\bm{\Theta},u}^{\rm{H}} \! \bm{W}_u \bm{G}_{\bm{\Theta},u} \bm{H}_{u,u}, \!\!\!\!\!\!&\text{for capacity max problem},\\ 
		\bm{H}_{u,u}^{\rm{H}}\bm{G}_{\bm{\Theta},u}^{\rm{H}}  \bm{G}_{\bm{\Theta},u} \bm{H}_{u,u}, &\text{for MSE min problem},
	\end{cases}\nonumber\\
 &\bm{\Pi}_u=\bm{F}_{\rm{D},u} \bm{F}_{\rm{D},u}^{\rm{H}},\\
  &\bm{B}_u=
	\begin{cases}
		\bm{H}_{u,u}^{\rm{H}} \bm{G}_{\bm{\Theta},u}^{\rm{H}} \bm{W}_u \bm{F}_{\rm{D},u}^{\rm{H}}, &\!\!\!\!\text{for capacity max problem},\\ 
		\bm{H}_{u,u}^{\rm{H}} \bm{G}_{\bm{\Theta},u}^{\rm{H}} \bm{F}_{\rm{D},u}^{\rm{H}}, &\!\!\!\!\text{for MSE min problem},
	\end{cases}\ \forall u.\nonumber
\end{align} 
Generally, the optimal $[\bm{X}_u]_{i,j}$'s can be obtained when the element-wise phase derivatives of the objective function of \textbf{Prob.10} w.r.t. $[\bm{X}_u]_{i,j}$'s equal zeros, i.e.,
\begin{align}\label{g_11}
    \underbrace{\Im\left\{ \big( [(\bm{\Phi}\bm{X}_u\bm{\Pi})^*]_{i,j}-[\bm{B}^*]_{i,j}\big)[\bm{X}_u]_{i,j}\right\}
  }_{\triangleq g_{i,j}(\bm{X}_u)}=0, \ \forall u,i, j,
\end{align} 
where $[(\bm{\Phi}\bm{X}_u\bm{\Pi})]_{i,j}$'s can be rewritten as
\begin{align}\label{factor_x_hyb_multi}
&[(\bm{\Phi}\bm{X}_u\bm{\Pi})]_{i,j}\nonumber=
{\rm{Tr}}\left(\bm{X}_u[\bm{\Pi}]_{:,j}[\bm{\Phi}]_{i,:}\right) \nonumber \\
=&[\bm{X}_u]_{i,j}\left[ ([\bm{\Pi}]_{:,j}[\bm{\Phi}]_{i,:})\right]_{j,i}\\
&+\underbrace{{\sum}_{m\not{=i}}{\sum}_{n\not{=}j}[\bm{X}_u]_{m,n}\left[ ([\bm{\Pi}]_{:,j}[\bm{\Phi}]_{i,:})\right]_{n,m}}_{\triangleq s_{u,i,j}}, \ \forall u,i, j.\nonumber
\end{align} 
By substituting (\ref{factor_x_hyb_multi}) into (\ref{g_11}), we have
\begin{subequations}
  \begin{align}\label{unmse_deri}
& \Im \left\{ \big( [(\bm{\Phi}\bm{X}_u\bm{\Pi})^*]_{i,j}-[\bm{B}^*]_{i,j}\big)[\bm{X}_u]_{i,j}\right\} \\
=&\Im \left\{ \left(s_{i,j}\!-\![\bm{B}]_{i,j}\right)^*[\bm{X}_u]_{i,j}\!+\!\left[ ([\bm{\Pi}]_{:,j}[\bm{\Phi}]_{i,:})\right]_{j,i}^*\right\}\!=\!0, \ \forall u,i, j.\label{unmse_deri_b}
\end{align}  
\end{subequations}
The last term $\left[ ([\bm{\Pi}]_{:,j}[\bm{\Phi}]_{i,:})\right]_{j,i}^*$ is a real scalar since it satisfies
$\left[ ([\bm{\Pi}]_{:,j}[\bm{\Phi}]_{i,:})\right]_{j,i}^*\!=\![\bm{\Pi}]_{j,j}[\bm{\Phi}]_{i,i}\!=\!\left([\bm{\Pi}]_{j,j}[\bm{\Phi}]_{i,i}\right)^*$.
Thus, the optimal $[\bm{\Theta}_u]_{i,j}$'s satisfying (\ref{unmse_deri_b}) for \textbf{Prob.10} are given by
\begin{align}\label{unmse_solu}
[\bm{\Theta}_u]_{i,j}=&{\rm{Phase}}\left\{s_{i,j}-[\bm{B}]_{i,j}\right\}
\nonumber\\ \text{or} \
&\pi +{\rm{Phase}}\left\{s_{i,j}-[\bm{B}]_{i,j}\right\}
, \ \forall u, i, j.
\end{align}

\subsubsection{Fully-Passive IRS-aided MIMO System}
In the fully-passive IRS-aided point-to-point MIMO system, the received signal at the user can be written as
\begin{align}\label{irs_channel}
  \bm{y}=\bm{H} \bm{s}+\bm{n}
  =\left(\bm{H}_0+\bm{H}_1\bm{\Lambda}_{\bm{\Theta}}\bm{H}_2\right) \bm{s}
 +\bm{n} ,
\end{align}
where $\bm{H}_0\in \mathbb{C}^{N_r \times N_t}$, $\bm{H}_1\in \mathbb{C}^{N_r \times K}$ and $\bm{H}_2\in \mathbb{C}^{K \times N_t}$  represent the BS-user direct channel, the IRS-user channel and the BS-IRS channel, respectively.
$\bm{\Lambda}_{\bm{\Theta}}\in \mathbb{C}^{K \times K}$ is the diagonal IRS reflection matrix subject to both diagonal structure constraints and constant modulus constraints\cite{TQiao_Tcom_IRS}.
The phase shift vector $\bm{\theta}$ corresponding to the IRS reflection matrix $\bm{\Lambda}_{\bm{\Theta}}$ is then defined as
\begin{align}
[\bm{\theta}]_{i}=\theta_{i}, \ 
\left[\bm{\Lambda}_{\bm{\Theta}}\right]_{i,i}=e^{ \jmath \theta_i}, \ \forall i.
\end{align}
Firstly, we consider the classical capacity maximization problem in the fully-passive IRS-aided MIMO system as follows:
\begin{align}
\textbf{Prob.11:} \ \max_{\bm{\Lambda}_{\bm{\Theta}}} \ & 
\log \left|\bm{\Sigma}^{-1}\bm{H}\bm{H}^{\rm{H}}+\bm{I}_{N_r}\right| \nonumber \\
{\rm{s.t.}} \ & |[\bm{\Lambda}_{\bm{\Theta}}]_{i,i}|=1, \ \forall i.
\end{align}
By leveraging the KKT optimality conditions, the element-wise phase derivatives of the objective function of \textbf{Prob.11} w.r.t. $\left[ \bm{\Lambda}_{\bm{\Theta}}\right] _{i,i}$'s equal zeros at the optimal $\left[ \bm{\Lambda}_{\bm{\Theta}}\right] _{i,i}$'s, i.e.,
\begin{align}\label{g_15}
 &\underbrace{\Im \!\left\{ \!\left[\bm{H}_2\bm{H}^{\rm{H}}\bm{\Sigma}^{-1}\!\!\left( \bm{\Sigma}^{-1}\! \bm{H}\bm{H}^{\rm{H}}\! +\!\bm{I}_{N_r} \right)^{-1}\!\!\bm{H}_1\right]_{i,i}\!\left[ \bm{\Lambda}_{\bm{\Theta}}\right]_{i,i}\! \right\}
  }_{\triangleq g_{i,i}(\bm{\Lambda}_{\bm{\Theta}})}
 \! \!=\!0, \ \forall i.
\end{align}
The left-hand side of (\ref{g_15}) can be further rewritten as
\begin{align}\label{irs_tr_deri}
& \Im \left\{ \left[\bm{H}_2\bm{H}^{\rm{H}}\bm{\Sigma}^{-1} \left( \bm{\Sigma}^{-1}\!\bm{H}\bm{H}^{\rm{H}} \!+\!\bm{I}_{N_r} \right)^{-1}\!\!\bm{H}_1\right]_{i,i}\left[ \bm{\Lambda}_{\bm{\Theta}}\right]_{i,i} \right\}\nonumber \\
=& \Im \left\{ \left[\bm{H}_2\bm{H}^{\rm{H}}\bm{\Sigma}^{-\frac{1}{2}} \left(\bm{\Sigma}^{-\frac{1}{2}}\bm{H}\bm{H}^{\rm{H}}\bm{\Sigma}^{-\frac{1}{2}} +\bm{I}_{N_r} \right)^{-1}\right.\right.\nonumber\\
&\left.\left.\quad \times \bm{\Sigma}^{-\frac{1}{2}}\bm{H}_1\right]_{i,i}\left[ \bm{\Lambda}_{\bm{\Theta}}\right]_{i,i} \right\}\nonumber \\
\overset{(g_1)}{=}& \Im \left\{ {\rm{Tr}}\left[ \bm{\Gamma}_i \left(\bm{M}_i^{\rm{H}}+e^{-\jmath \theta_i}\bm{\Gamma}_i^{\rm{H}}\right) \left(\bm{M}_i\bm{M}_i^{\rm{H}}+e^{-\jmath \theta_i}\bm{M}_i\bm{\Gamma}_i^{\rm{H}}\right.\right.\right.\nonumber\\
&\left.\left.\left.\quad+e^{\jmath \theta_i}\bm{\Gamma}_i\bm{M}_i^{\rm{H}}+\bm{\Gamma}_i\bm{\Gamma}_i^{\rm{H}}
+\bm{I}_{N_r}\right)^{-1} \right]e^{\jmath \theta_i} \right\}\nonumber \\
\overset{(g_2)}{=}& \Im\! \left\{\! {\rm{Tr}}\!\left[ e^{\jmath \theta_i}\bm{\Gamma}_i\bm{M}_i^{\rm{H}} \!
\left(\bm{\Phi}_i\!+\!e^{-\jmath \theta_i}\bm{M}_i\bm{\Gamma}_i^{\rm{H}}
e^{\jmath \theta_i}\bm{\Gamma}_i\bm{M}_i^{\rm{H}}\right)^{-1} \!\right]\!+\!c_i\!\right\}\nonumber\\
\overset{(g_3)}{=}& \Im \left\{ {\rm{Tr}}\left[ e^{\jmath \theta_i}\bm{u}_i\bm{v}_i^{\rm{H}} 
\left(\bm{\Psi}_i -  \bm{a}_i \bm{a}_i^{\rm{H}}\right)^{-1} \right]+c_i\right\}\nonumber\\
\overset{(g_4)}{=}& \Im \left\{e^{\jmath \theta_i} \bm{v}_i^{\rm{H}} 
\left(\bm{\Psi}_i^{-1}\!\!+\!\frac{\bm{\Psi}_{i}^{-1}\bm{a}_i \bm{a}_i^{\rm{H}}\bm{\Psi}_{i}^{-1}}
{1-\bm{a}_i^{\rm{H}}\bm{\Psi}_{i}^{-1}\bm{a}_i}\right)  
\bm{u}_i \! +\!c_i\right\}, \ \forall i,
\end{align} 
where $(g_1)$ holds by rewriting $\bm{\Sigma}^{-\frac{1}{2} }\bm{H}$ in terms of $\left[ \bm{\Lambda}_{\bm{\Theta}}\right]_{i,i}$ as 
\begin{align}\label{auxiliary_irs_channel}
  \bm{\Sigma}^{-\frac{1}{2} }\bm{H}
  \!=\!&\left[ \bm{\Lambda}_{\bm{\Theta}}\right]_{i,i}\bm{\Gamma}_i
  \!+\!\underbrace{\bm{\Sigma}^{-\frac{1}{2} }\bm{H}_0\!+\!{\sum}_{n\not{=}i} \left[ \bm{\Lambda}_{\bm{\Theta}}\right]_{n,n}\bm{\Gamma}_n }_{\triangleq \bm{M}_i},
\end{align}
where $\bm{\Gamma}_i=[\bm{\Sigma}^{-\frac{1}{2} } \bm{H}_1]_{:,i}[\bm{H}_2]_{i,:}$.
The equality $(g_2)$ in (\ref{irs_tr_deri}) is obtained by defining $\bm{\Phi}_i=\bm{M}_i\bm{M}_i^{\rm{H}}+\bm{\Gamma}_i\bm{\Gamma}_i^{\rm{H}}+\bm{I}_{N_r}$,
 $c_i={\rm{Tr}}\left[ \bm{\Gamma}_i\bm{\Gamma}_i^{\rm{H}} 
 \left(\bm{\Phi}_i+e^{-\jmath \theta_i}\bm{M}_i\bm{\Gamma}_i^{\rm{H}}+
 e^{\jmath \theta_i}\bm{\Gamma}_i\bm{M}_i^{\rm{H}}\right)^{-1} \right]$, where $c_i$ is a real scalar independent of the optimal $\left[ \bm{\Lambda}_{\bm{\Theta}}\right] _{i,i}$.
 The equality $(g_3)$ holds based on 
$\bm{a}_i=e^{\jmath \theta_i} \bm{v}_i - \bm{u}_i,
 \bm{\Psi}_i = \bm{\Phi}_i + \bm{v}_i \bm{v}_i^{\rm{H}} + \bm{u}_i \bm{u}_i^{\rm{H}}
$,
where $\bm{v}_i$ and $\bm{u}_i$ come from the SVD of the rank-one matrix $\bm{M}_i\bm{\Gamma}_i^{\rm{H}}$,
i.e., $\bm{M}_i\bm{\Gamma}_i^{\rm{H}}=\bm{v}_i\bm{u}_i^{\rm{H}}$.
The equality $(g_4)$ holds similarly to $(a_2)$ in (\ref{left_diag_log}).
Then, by substituting (\ref{irs_tr_deri}) into (\ref{g_15}), we have
\begin{subequations}\label{g_log}
\begin{align}
  & \Im\! \left\{\!\! \left[\!\bm{H}_2\bm{H}^{\rm{H}}\bm{\Sigma}^{-1}\!\!\left( \bm{\Sigma}^{-1} \!\bm{H}\bm{H}^{\rm{H}} \!+\!\bm{I}_{N_r} \right)^{-1}\!\!\bm{H}_1\!\right]_{i,i}\!\left[ \bm{\Lambda}_{\bm{\Theta}}\right]_{i,i} \!\right\}  \\
  =&\Im \left\{\frac{A_{2,i} e^{\jmath \theta_i}+C_{2,i} }
  {D_{2,i}} 
 +c_i\right\}=0, \ \forall i,\label{g_log_d}
\end{align}
\end{subequations}
where \vspace{-2mm}
\begin{align}
  & A_{2,i} =\bm{v}_i^{\rm{H}} \bm{\Psi}_i^{-1} \bm{u}_i,\\
  & C_{2,i} = \bm{v}_i^{\rm{H}} \bm{\Psi}_i^{-1} \bm{u}_i  \bm{u}_i^{\rm{H}} \bm{\Psi}_i^{-1} \bm{v}_i
  - \bm{v}_i^{\rm{H}} \bm{\Psi}_i^{-1} \bm{v}_i\bm{u}_i^{\rm{H}} \bm{\Psi}_i^{-1} \bm{u}_i,\nonumber\\
 & D_{2,i}\!=\!2{\rm{Re}}\left\{\bm{v}_i^{\rm{H}} \bm{\Psi}_i^{-1} \bm{u}_i e^{\jmath \theta_i} \right\}  \!+\! 1\!-\!\bm{v}_i^{\rm{H}} \bm{\Psi}_i^{-1} \bm{v}_i\!-\!\bm{u}_i^{\rm{H}} \bm{\Psi}_i^{-1} \bm{u}_i, \ \forall i.\nonumber
\end{align}
Similar to (\ref{gen_hy_log_b}), since $C_{2,i}$'s, $D_{2,i}$'s and $c_i$'s are real scalars,
the optimal $\left[\bm{\theta}\right]_i$'s satisfying (\ref{g_log_d}) for \textbf{Prob.11} are given by
\begin{align}
\left[\bm{\theta}\right]_i={\rm{Phase}}\left\{
  A_{2,i}^* 
\right\} \ \text{or} \
\pi +{\rm{Phase}}\left\{
  A_{2,i}^* 
\right\}, \ \forall i.
\end{align}

Additionally, we consider the MSE minimization problem in the fully-passive IRS-aided point-to-point MIMO system, which is formulated as
\vspace{-2mm}
\begin{align}
\textbf{Prob.12:} \ \min_{\bm{\Lambda}_{\bm{\Theta}}} \ & 
{\rm{Tr}}\left[ \left(\bm{\Sigma}^{-1}\bm{H}\bm{H}^{\rm{H}}+\bm{I}_{N_r}\right)^{-1} \right]\nonumber \\
{\rm{s.t.}} \ & |[\bm{\Lambda}_{\bm{\Theta}}]_{i,i}|=1, \ \forall i.
\end{align}
Since the element-wise phase derivatives of the objective function of \textbf{Prob.12} w.r.t. $\left[ \bm{\Lambda}_{\bm{\Theta}}\right] _{i,i}$'s equal zeros at the optimal solution, we have
\vspace{-2mm}
\begin{align}\label{g_16}
\setlength{\belowdisplayskip}{1pt}
  \underbrace{\Im\! \left\{ \left[\bm{H}_2\bm{H}^{\rm{H}}\bm{\Sigma}^{-1}\!\!\left( \bm{\Sigma}^{-1}\! \bm{H}\bm{H}^{\rm{H}}\! +\!\bm{I}_{N_r} \right)^{-2}\!\!\bm{H}_1\right]_{i,i}\!\left[ \bm{\Lambda}_{\bm{\Theta}}\right]_{i,i}\! \right\}
  }_{\triangleq g_{i,i}(\bm{\Lambda}_{\bm{\Theta}})}
  \!=\!0, \ \forall i.
\end{align}
The left-hand side of (\ref{g_16}) can be further rewritten as
\begin{align}\label{g_tr}
&\Im \left\{ \left[\bm{H}_2\bm{H}^{\rm{H}}\bm{\Sigma}^{-1} \left( \bm{\Sigma}^{-1} \bm{H}\bm{H}^{\rm{H}} +\bm{I}_{N_r} \right)^{-2}\bm{H}_1\right]_{i,i}\left[ \bm{\Lambda}_{\bm{\Theta}}\right]_{i,i} \right\}\nonumber \\
\overset{(h)}{=}&\Im \left\{e^{\jmath \theta_i} \bm{v}_i^{\rm{H}} 
\left(\bm{\Psi}_i^{-1}+\frac{\bm{\Psi}_{i}^{-1}\bm{a}_i \bm{a}_i^{\rm{H}}\bm{\Psi}_{i}^{-1}}
{1-\bm{a}_i^{\rm{H}}\bm{\Psi}_{i}^{-1}\bm{a}_i}\right) \right. \nonumber\\
&\left.\qquad\times\left(\bm{\Psi}_i^{-1}+\frac{\bm{\Psi}_{i}^{-1}\bm{a}_i \bm{a}_i^{\rm{H}}\bm{\Psi}_{i}^{-1}}
{1-\bm{a}_i^{\rm{H}}\bm{\Psi}_{i}^{-1}\bm{a}_i}\right)
\bm{u}_i  +c_i\right\}\nonumber\\
=&\Im \left\{\frac{e^{\jmath \theta_i}\left(d_i \eta_{1,i} + d_i \eta_{2,i} + d_i^2 \varepsilon_i + b_{1,i} - r_{0,i} \lambda_i\right)  }
{\left(2{\rm{Re}}\left\{\lambda_i e^{\jmath \theta_i} \right\}  + d\right)^2 }\right.\nonumber\\
&\ +\frac{e^{-\jmath \theta_i}\left(  (\lambda_i^*)^2 \varepsilon_i +b_{2,i}- r_{0,i} \lambda_i^* \right) }
{\left(2{\rm{Re}}\left\{\lambda_i e^{\jmath \theta_i} \right\}  + d\right)^2 }\\
&  \left. \ +\frac{2 d_i \lambda_i^* \varepsilon_i + \lambda_i^* \eta_{1,i} + \lambda_i^* \eta_{2,i} - b_{0,i}
- r_{0,i} d_i}{\left(2{\rm{Re}}\left\{\lambda_i e^{\jmath \theta_i} \right\}  + d\right)^2} 
+c_i\right\}, \ \forall i,\nonumber
\end{align}
where $(h)$ holds similarly to (\ref{irs_tr_deri}) and  we  have
\begin{align}
  \lambda_i\!=&\!\bm{v}_i^{\rm{H}} \bm{\Psi}_i^{-1} \bm{u}_i,
\varepsilon_i \!=\! \bm{v}_i^{\rm{H}} \bm{\Psi}_i^{-2} \bm{u}_i, 
d_i\!=\! 1\!-\!\bm{v}_i^{\rm{H}} \bm{\Psi}_i^{-1} \!\bm{v}_i\!-\!\bm{u}_i^{\rm{H}} \bm{\Psi}_i^{-1}\! \bm{u}_i,\nonumber\\
\eta_{1,i}&= \bm{v}_i^{\rm{H}} \bm{\Psi}_i^{-2} \left(\bm{u}_i \bm{u}_i^{\rm{H}}+\bm{v}_i \bm{v}_i^{\rm{H}} \right) \bm{\Psi}_i^{-1} \bm{u}_i,\nonumber\\
  \eta_{2,i}&= \bm{v}_i^{\rm{H}} \bm{\Psi}_i^{-1} \left(\bm{u}_i \bm{u}_i^{\rm{H}}+\bm{v}_i \bm{v}_i^{\rm{H}} \right) \bm{\Psi}_i^{-2} \bm{u}_i,\nonumber\\
  b_{0,i}&=\bm{v}_i^{\rm{H}} \bm{\Psi}_i^{-1} \left(\bm{u}_i \bm{u}_i^{\rm{H}}+\bm{v}_i \bm{v}_i^{\rm{H}} \right) 
  \bm{\Psi}_i^{-2} \bm{v}_i  \bm{u}_i^{\rm{H}} \bm{\Psi}_i^{-1} \bm{u}_i\nonumber\\
  &~~+\bm{v}_i^{\rm{H}} \bm{\Psi}_i^{-1} \bm{v}_i  \bm{u}_i^{\rm{H}} \bm{\Psi}_i^{-2}
  \left(\bm{u}_i \bm{u}_i^{\rm{H}}+\bm{v}_i \bm{v}_i^{\rm{H}} \right) \bm{\Psi}_i^{-1}\bm{u}_i
  ,\nonumber\\
  b_{1,i}&=\bm{v}_i^{\rm{H}} \bm{\Psi}_i^{-1} \left(\bm{u}_i \bm{u}_i^{\rm{H}}+\bm{v}_i \bm{v}_i^{\rm{H}} \right) 
  \bm{\Psi}_i^{-2} \left(\bm{u}_i \bm{u}_i^{\rm{H}}+\bm{v}_i \bm{v}_i^{\rm{H}} \right) \bm{\Psi}_i^{-1} \bm{u}_i
\nonumber\\
  \!+&\bm{v}_i^{\rm{H}} \bm{\Psi}_i^{-1} \left(\bm{\Psi}_i^{-1} \bm{u}_i  \bm{v}_i^{\rm{H}} \bm{\Psi}_i^{-2}\bm{v}_i^{\rm{H}}\bm{u}_i 
  \!+\!\bm{v}_i  \bm{u}_i^{\rm{H}} \bm{\Psi}_i^{-2} 
  \bm{u}_i^{\rm{H}} \bm{v}_i \right)
  \bm{\Psi}_i^{-1} \bm{u}_i
  ,\nonumber\\
  b_{2,i}&=\bm{v}_i^{\rm{H}} \bm{\Psi}_i^{-1} \bm{v}_i 
   \bm{u}_i^{\rm{H}} \bm{\Psi}_i^{-2} \bm{v}_i
   \bm{u}_i^{\rm{H}} \bm{\Psi}_i^{-1} \bm{u}_i,\\
  r_{0,i}&=\bm{v}_i^{\rm{H}} \bm{\Psi}_i^{-1} \bm{v}_i 
   \bm{u}_i^{\rm{H}} \bm{\Psi}_i^{-2} \bm{u}_i
  +\bm{v}_i^{\rm{H}} \bm{\Psi}_i^{-2} \bm{v}_i 
   \bm{u}_i^{\rm{H}} \bm{\Psi}_i^{-1} \bm{u}_i,~~\forall i.\nonumber
  \end{align}
By substituting (\ref{g_tr}) into (\ref{g_16}), we have
\begin{subequations}
\begin{align}
  &\Im\! \left\{\!\! \left[\!\bm{H}_2\bm{H}^{\rm{H}}\bm{\Sigma}^{-1}\!\!\left( \bm{\Sigma}^{-1} \!\bm{H}\bm{H}^{\rm{H}} \!+\!\bm{I}_{N_r} \right)^{-2}\!\!\bm{H}_1\!\right]_{i,i}\!\left[ \bm{\Lambda}_{\bm{\Theta}}\right]_{i,i} \!\right\}\\
 =&\Im\left\{ \frac{A_{3,i} e^{\jmath \theta_i} + B_{3,i} e^{-\jmath \theta_i} + C_{3,i}}
  {D_{3,i}}+c_i \right\}=0, \ \forall i ,\label{gen_irs_tr}
\end{align}
\end{subequations}
where\vspace{-2mm}
\begin{align}
  A_{3,i}=&d_i \eta_{1,i} + d_i \eta_{2,i} + d_i^2 \varepsilon_i + b_{1,i} - r_{0,i} \lambda_i,\nonumber\\
  B_{3,i}=& (\lambda_i^*)^2 \varepsilon_i +b_{2,i}- r_{0,i} \lambda_i^*,\nonumber\\
  C_{3,i}=& 2 d_i \lambda_i^* \varepsilon_i + \lambda_i^* \eta_{1,i} + \lambda_i^* \eta_{2,i} - b_{0,i}
  - r_{0,i} d_i,\nonumber\\
  D_{3,i}=& \left(2{\rm{Re}}\left\{\lambda_i e^{\jmath \theta_i} \right\}  + d_i\right)^2,\ \forall i.
\end{align}
It is noted that (\ref{gen_irs_tr}) has the same form as (\ref{gen_hy_tr_c}) for \textbf{Prob.9}. 
Thus, the optimal $\left[\bm{\theta}\right]_i$'s to \textbf{Prob.12} can be obtained similarly.

Moreover, like \textbf{Prob.10}, when the fully-passive IRS-aided MU-MIMO system is taken into account,
the general WMMSE minimization problem is formulated as
\vspace{-1mm}
\begin{align}
\textbf{Prob.13:} \ \min_{\bm{\Lambda}_{\bm{\Theta}}} \ &  {\rm{Tr}}(\bm{\Phi}\bm{\Lambda}_{\bm{\Theta}}\bm{\Pi}\bm{\Lambda}_{\bm{\Theta}}^{\rm{H}})\!-\!{\rm{Tr}}(\bm{B}^{\rm{H}}\bm{\Lambda}_{\bm{\Theta}})\!-\!{\rm{Tr}}(\bm{B}\bm{\Lambda}_{\bm{\Theta}}^{\rm{H}}) \nonumber \\
{\rm{s.t.}}  \ & |[\bm{\Lambda}_{\bm{\Theta}}]_{i,i}|=1, \ \forall i.
\end{align}
Since the element-wise phase derivatives of the objective function w.r.t. $\left[ \bm{\Lambda}_{\bm{\Theta}}\right] _{i,i}$'s are zeros at the optimal solution, i.e.,
\begin{align}\label{g_14}
 \underbrace{\Im\!\left\{ \!\left[\left(\bm{\Phi}\bm{\Lambda}_{\bm{\Theta}}\bm{\Pi}\right)^*\right]_{i,i}\!\left[\bm{\Lambda}_{\bm{\Theta}}\right]_{i,i} \! \right\}\!-\!\!\Im\!\left\{ \!\left[\bm{B}^*\right]_{i,i}\!\left[\bm{\Lambda}_{\bm{\Theta}}\right]_{i,i} \! \right\}
  }_{\triangleq g_{i,i}(\bm{\Lambda}_{\bm{\Theta}})}\!\!=\!0, \ \forall i,
\end{align} 
the optimal solution for \textbf{Prob.13} can be easily derived as
\vspace{-1mm}
\begin{align}
\left[\bm{\theta}\right]_i=&{\rm{Phase}}\left\{\left[\bm{\Phi}\bm{\Lambda}_{\bm{\Theta}}\bm{\Pi}\right]_{i,i} +\left[\bm{B}\right]_{i,i}\right\}
\nonumber\\ \text{or} \
&\pi +{\rm{Phase}}\left\{\left[\bm{\Phi}\bm{\Lambda}_{\bm{\Theta}}\bm{\Pi}\right]_{i,i} +\left[\bm{B}\right]_{i,i}\right\}, \ \forall i.
\end{align}

In a nutshell, a series of optimization problems in the wireless systems associated with constant modulus constraints are investigated in this subsection,
whose optimal solutions are available using the proposed element-wise phase derivatives.

\vspace{-3mm}
\subsection{A Novel AO Algorithm}
It  follows from  Sec. \ref{sec_const_app} that  these element-wise phase derivatives associated with \textbf{Prob.8}$\sim$\textbf{Prob.13}  can be mainly classified into two  forms, 
and each form always has two zero-derivative points,
as  summarized in the following proposition.

\noindent \textbf{Proposition 1.}
\textit{For different types of objective functions, the element-wise phase derivatives under constant modulus constraints
can be mainly summarized as the following two general forms, 
i.e., the linear form and the conjugate linear form,
which are shown as
\begin{subequations}
  \begin{numcases}
    {g_{i,j}(\bm{X})\!=\!\!}
  \!\Im \!\left\{A^{\rm{L}}_{i,j}[\bm{X}]_{i,j}\!+\!C^{\rm{L}}_{i,j}\right\},  \forall i, j,  \text{for trace-linear,}\nonumber\\ 
    \text{trace-quadratic, log-determinant functions},\label{linear_form}\\
  \Im\! \left\{A^{\rm{CL}}_{i,j}[\bm{X}]_{i,j}\!+B^{\rm{CL}}_{i,j}[\bm{X}]^*_{i,j}+C^{\rm{CL}}_{i,j}\right\},  \forall i, j,\nonumber \\ 
  \!\qquad \qquad \qquad \quad \text{for trace-inverse function},\label{conj_linear_form}
  \end{numcases} 
\end{subequations} 
where $A^{\rm{L}}_{i,j}$'s, $A^{\rm{CL}}_{i,j}$'s, $B^{\rm{CL}}_{i,j}$'s and $C^{\rm{CL}}_{i,j}$'s are all complex scalars and $C^{\rm{L}}_{i,j}$'s are real scalars.
In particular, the linear element-wise phase derivative in (\ref{linear_form}) can be regarded as a simplified case of its conjugate linear counterpart
by setting $B^{\rm{CL}}_{i,j}=0$ and $\Im \left\{ C^{\rm{CL}}_{i,j}\right\}=0,  \forall i, j$ in (\ref{conj_linear_form}).
Moreover, there are two points $\left\{\left[\bm{X}_{1}\right]_{i,j},\left[\bm{X}_{2}\right]_{i,j}\right\}$ satisfying $g_{i,j}(\bm{X})=0, \forall i,j$, 
from which the  optimal solution  can be  determined as
\begin{align}\label{opt_judge}
  & \left[\bm{X}_{\rm{opt}}\right]_{i,j}= 
   \arg_{\left\{\left[\bm{X}_{1}\right]_{i,j},\left[\bm{X}_{2}\right]_{i,j}\right\}} \\
  &\!\!\!  \begin{cases}
      \!\!  \Re\!\left\{\! A^{\rm{CL}}_{i,j}[\bm{X}]_{i,j}\!-\!B^{\rm{CL}}_{i,j}[\bm{X}]^*_{i,j} \!\right\}\!\!\geq\! 0,
   \!\!\! \!& \text { for min problem} , \\
  \!\!\Re\!\left\{\! A^{\rm{CL}}_{i,j}[\bm{X}]_{i,j}\!-\!B^{\rm{CL}}_{i,j}[\bm{X}]^*_{i,j} \!\right\}\!\!<\! 0,
  \!\!\! \!& \text { for max problem},
		\end{cases}\!\forall i, j.\nonumber
  \end{align}}

\noindent\textit{Proof.}
The detailed proof is shown in Appendix \ref{proof_gen_form}.

Based on \textbf{proposition 1},  
it is seen that the optimal $[\bm{X}]_{i,j}$'s are obtained by aligning their phase-shifts with the  corresponding  counterparts
jointly determined by $A^{\rm{CL}}_{i,j}$'s, $B^{\rm{CL}}_{i,j}$'s and $C^{\rm{CL}}_{i,j}$'s,
which are all related to $[\bm{X}]_{m,n}$'s, $m \neq i, n \neq j$ and need to be frequently calculated in each iteration of updating $[\bm{X}]_{i,j}$'s.
In addition, the calculations of $A^{\rm{CL}}_{i,j}$'s, $B^{\rm{CL}}_{i,j}$'s and $C^{\rm{CL}}_{i,j}$'s all involve complicated matrix inversion with complexity of $\mathcal{O}\left(N_r^3\right)$ and SVD with complexity of $\mathcal{O}\left(2N_r N_t^2 + N_r^3\right)$.
It is evident that this complexity will become enormous as $N_r$ and $N_t$ increases.
In order to avoid the frequent calculations of $A^{\rm{CL}}_{i,j}$'s, $B^{\rm{CL}}_{i,j}$'s and $C^{\rm{CL}}_{i,j}$'s,
we next derive the optimal solution directly based on the functions $g_{i,j}(\bm{X})$'s associated with the original element-wise phase derivatives, 
which is shown in \textbf{Proposition 2}.

\setlength{\textfloatsep}{0.2cm}
\begin{algorithm}[tp!]
  \linespread{0.78}
  \caption{A Novel AO Algorithm for Solving Problems under Constant Modulus Constraints}	\label{EE-WF-SP} 
\begin{algorithmic}[1]

\REQUIRE Arbitrary five feasible solutions $\bm{\widehat X}_m^{(0)}$, $ m = 1,\cdots,5$; iteration index $t=0$;
convergence threshold $\epsilon $.
\REPEAT
\FOR{$ i = 1 $ to $ N_t $, $ j = 1 $ to $ N_{rf} $}
\STATE Calculate $ g_{i,j}(\bm{\widehat X}_m^{(t)}), m = 1,\cdots, 5 $ and the auxiliary vector $\bm{w}^{(t)}_{i,j}$ as in (\ref{general_equ}).
\STATE Update $\left[\bm{\Theta}^{(t)}\right]_{i,j}$ as in (\ref{opt_gen})
and obtain $ [\bm{ X}^{(t)}]_{i,j} $ according to (\ref{def_cons}).

\ENDFOR

\STATE Update $\bm{\widehat X}_m^{(t+1)}=\bm{ X}^{(t)} $ for an arbitrary $m \in [1,5]$.
\STATE $ t= t+1 $. 
\UNTIL{The increment/decrement of the objective function value between two consecutive iterations is less than $\epsilon$.} 
\RETURN $ \bm{X}$. 
\end{algorithmic}
\end{algorithm}
\setlength{\floatsep}{0.2cm}

\noindent \textbf{Proposition 2.}
\textit{Define arbitrary five feasible solutions satisfying constant modulus constraints, i.e., $\bm{\widehat X}_m$, $ m = 1,\cdots,5$ and calculate their corresponding $g_{i,j}(\bm{\widehat X}_m)$'s,
we have
\vspace{-0.75mm}
\begin{align}\label{general_equ}
    \bm{w}_{i,j} \!=\!\bm{R}_{i,j}^{-1} \bm{t}_{i,j} \text{ with } \bm{R}_{i,j} \!\in \!\mathbb{C}^{5 \times 5}, \bm{t}_{i,j}\! \in\! \mathbb{C}^{5 \times 1},  \  \forall i,j,
\end{align}
where 
\vspace{-2.1mm}
\begin{align}
     &[\bm{R}_{i,j}]_{m,:} =\bigg[ \Im \{[\bm{\widehat X}_m]_{i,j} \}\!-\![\bm{t}_{i,j}]_m\Re \{ [\bm{\widehat X}_m]_{i,j}\},
     \Re \{[\bm{\widehat X}_m]_{i,j} \}\!\nonumber\\
   &~~+\![\bm{t}_{i,j}]_m \Im\{ [\bm{\widehat X}_m]_{i,j}\}, -\Im \{[\bm{\widehat X}_m]_{i,j} \}\!-\![\bm{t}_{i,j}]_m\Re \{ [\bm{\widehat X}_m]_{i,j}\}, \nonumber\\
   &~~\Re \{[\bm{\widehat X}_m]_{i,j} \}\!-\![\bm{t}_{i,j}]_m \Im\{ [\bm{\widehat X}_m]_{i,j}\},
   1
   \bigg] ,\nonumber\\
  &[\bm{t}_{i,j}]_m=\tan ( \angle g_{i,j}(\bm{\widehat X}_m) ),\ \forall i,j,m .
\end{align}
Then, we  obtain the optimal solutions of optimization problems with conjugate linear element-wise phase derivatives  as 
\vspace{-1mm}
\begin{align}\label{opt_gen}
	\left[\bm{\Theta}\right]_{i,j}\!= \!
    \begin{cases}
		-\arctan \left(\frac{\widehat{z}_{2,i,j}}{\widehat{z}_{1,i,j}}\right)+\arcsin \left(-\frac{[\bm{w}_{i,j}]_5}{\sqrt{\widehat{z}_{1,i,j}^2+\widehat{z}_{2,i,j}^2}}\right) 
   \\ \text{ or }  
  \pi\!-\!\arctan\! \left(\frac{\widehat{z}_{2,i,j}}{\widehat{z}_{1,i,j}}\right) \!+\!\arcsin\!\left(\!-\frac{[\bm{w}_{i,j}]_5}{\sqrt{\widehat{z}_{1,i,j}^2+\widehat{z}_{2,i,j}^2}}\right),\\
  \qquad \qquad \qquad \qquad \qquad \qquad \ \text { if } \widehat{z}_{1,i,j} \geq 0,\ \forall i,j, \\
	-\pi+\arctan \left(\frac{\widehat{z}_{2,i,j}}{\widehat{z}_{1,i,j}}\right)+\arcsin \left(-\frac{[\bm{w}_{i,j}]_5}{\sqrt{\widehat{z}_{1,i,j}^2+\widehat{z}_{2,i,j}^2}}\right) 
  \\ \text{ or }  
  \arctan \left(\frac{\widehat{z}_{2,i,j}}{\widehat{z}_{1,i,j}}\right) +\arcsin\left(-\frac{[\bm{w}_{i,j}]_5}{\sqrt{\widehat{z}_{1,i,j}^2+\widehat{z}_{2,i,j}^2}}\right), \\
	\qquad \qquad \qquad \qquad \qquad \qquad \  \text { if } \widehat{z}_{1,i,j}<0,\ \forall i,j,
		\end{cases}
\end{align} where $\widehat{z}_{1,i,j}=[\bm{w}_{i,j}]_1-[\bm{w}_{i,j}]_3$ and
$\widehat{z}_{2,i,j}=[\bm{w}_{i,j}]_2+[\bm{w}_{i,j}]_4, \forall i,j$.
In particular, for the special case of the linear element-wise phase derivatives, we have $\bm{w}_{i,j}\!=\![ [\bm{w}_{i,j}]_1, [\bm{w}_{i,j}]_2, 0,0,0 ]^{\rm{T}}$ 
and $\widehat{z}_{1,i,j}$ does not  affect the optimal $\left[\bm{\Theta}\right]_{i,j}$.}

\noindent\textit{Proof.}
The detailed proof is shown in Appendix \ref{proof_equ_method}.

  \begin{figure*}[t]
    \vspace{-12mm}
      \centering
      \subfigure[ Uplink MU-SIMO system]{
  \includegraphics[width =0.31\textwidth]{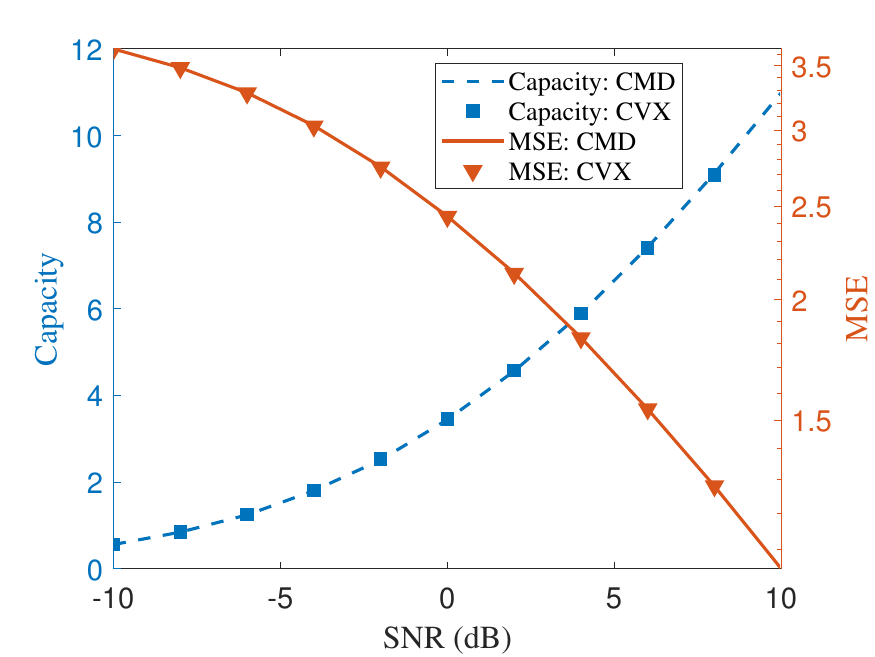}
  \label{fig_sim_diag1}
      }
      \subfigure[ Amplitude-adjustable IRS-aided MIMO system]{
  \includegraphics[width =0.31\textwidth]{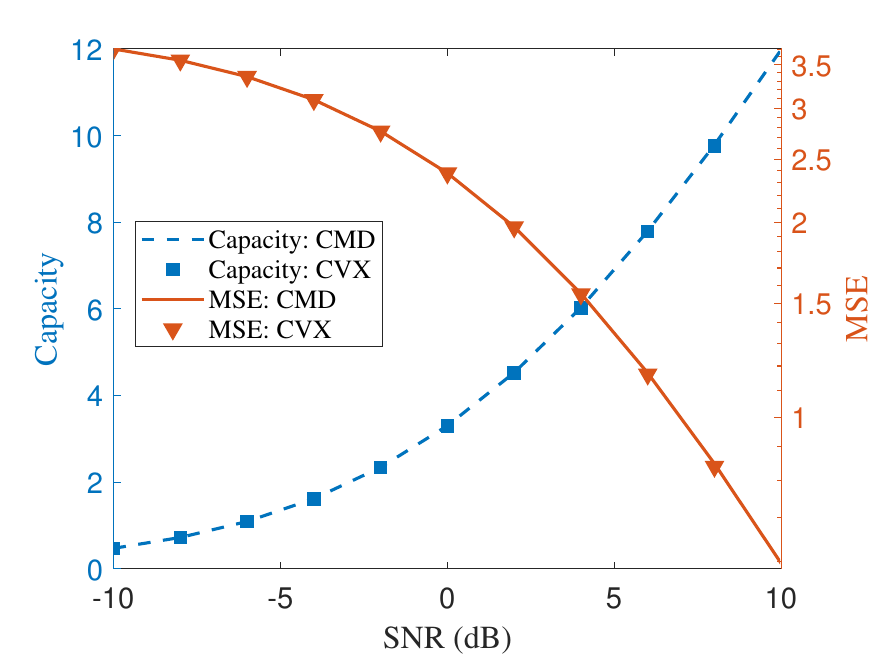}
  \label{fig_sim_diag2}
      }%
      \vspace{-2mm}
      \caption{The capacity and MSE performance comparison in the two  systems considered in Sec. \ref{sec_diag}.}
      \vspace{-4mm}
      \label{fig_diag}
  \end{figure*}
\begin{figure*}[t]
		\centering
    \subfigure[For different feasible solutions]
    {
    \includegraphics[width=0.31\textwidth]{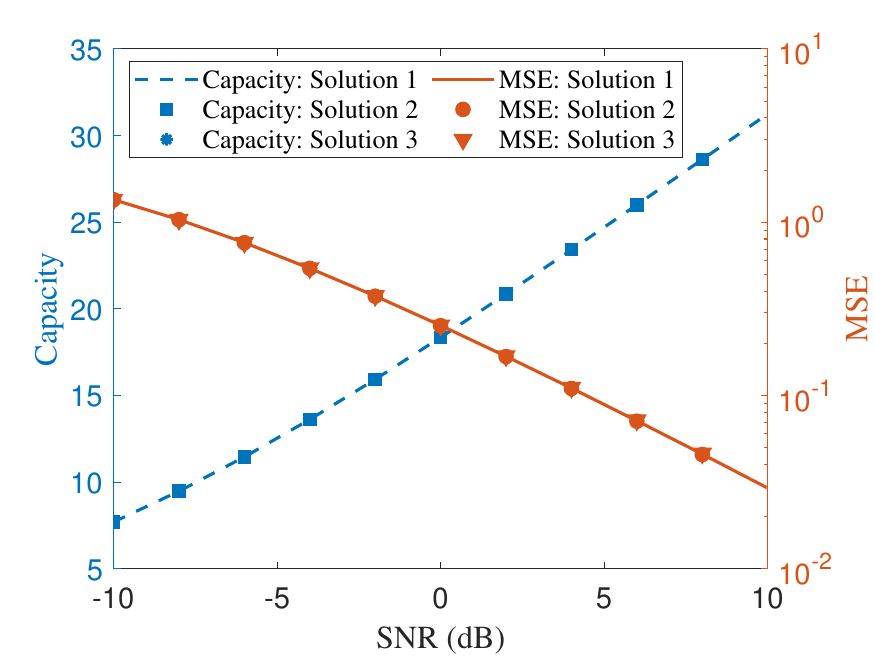} 
   \label{fig_sim_feasi_point}
    }
    \subfigure[For all studied algorithms]
    {
    \includegraphics[width=0.31\textwidth]{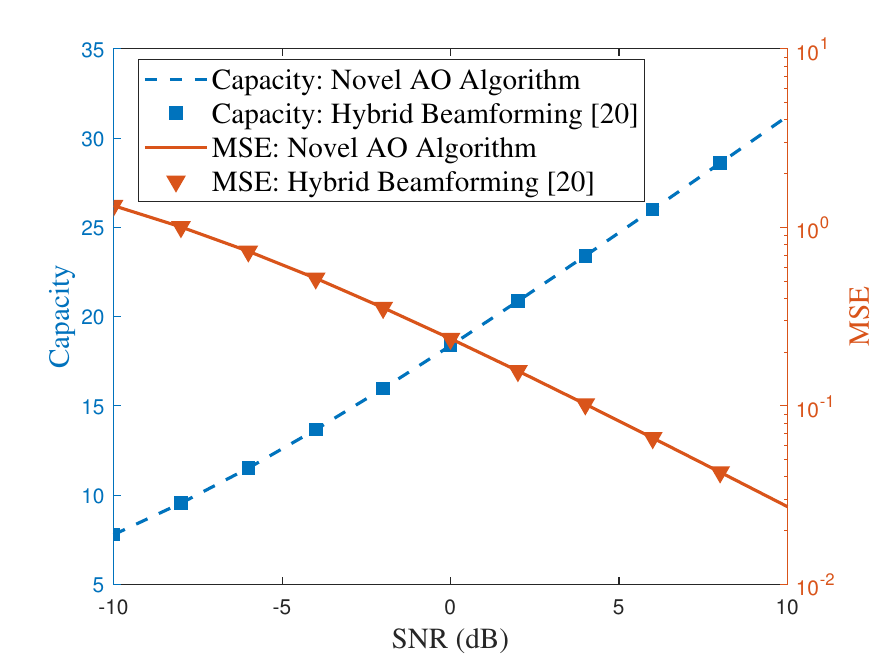} 
   \label{fig_sim_hybrid}
    }
    \subfigure[Convergence behaviors of all studied algorithms]
    {
    \includegraphics[width=0.31\textwidth]{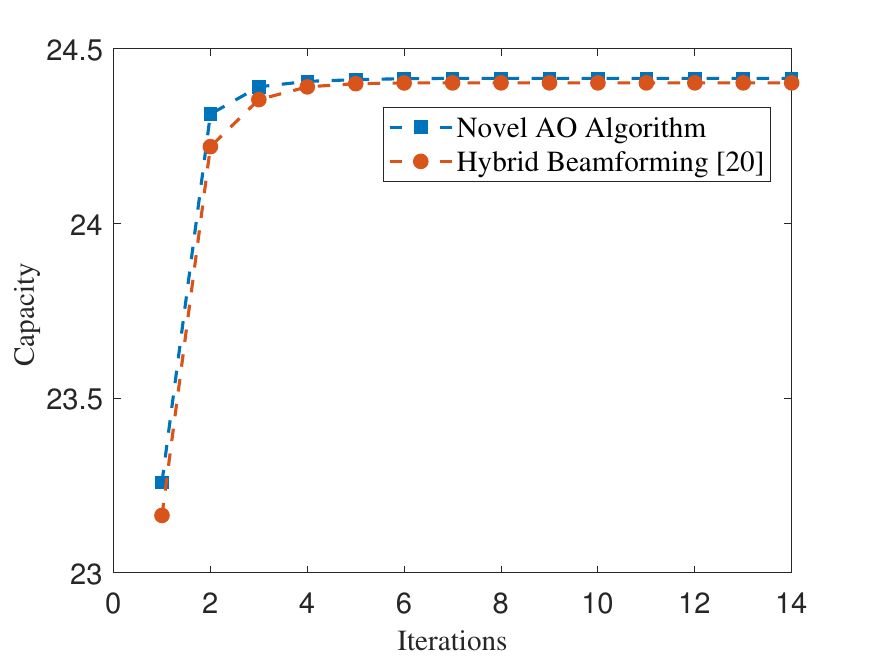} 
   \label{fig_sim_conver}
    }%
    \vspace{-2mm}
    \caption{The capacity and MSE performance comparison in the hybrid analog-digital MIMO system.}
    \label{fig_feasi_hybrid}
    \vspace{-5mm}
\end{figure*}

Based on \textbf{Proposition 2}, we next aim to develop a novel AO algorithm with the aid of five arbitrary feasible solutions to determine the optimal solutions of  all above optimization problems under constant modulus constraints, 
which is summarized in \textbf{Algorithm 1}.

\noindent\textit{Remark 2:}
 In  fact,  the  element-wise phase derivatives for different optimization problems with  constant modulus constraints  can  be roughly summarized as a  general  (conjugate) linear form, based on which a  novel AO algorithm with the  advantages of low complexity and guaranteed performance is  developed.

 \vspace{-3mm}

\section{Simulations and Discussions}\label{sec_sim}
In this section, numerical simulation results are provided to evaluate 
the performance of the derived optimal closed-form solutions based on
complex matrix derivatives in Sec. \ref{sec_diag_app} (also referred to as CMD-based algorithm),
and the novel AO algorithm in \textbf{Algorithm 1},
which are respectively proposed for tackling the optimization problems under diagonal structure constraints
and constant modulus constraints.
\vspace{-3.5mm}
\subsection{Diagonal Structure Constraints}
\vspace{-1mm}
We firstly consider the uplink MU-SIMO system, where $K=4$ single-antenna users transmit signals to
the BS equipped with $N_t=6$ antennas.
Moreover, the maximum transmit power of each user is assumed as $P_1 = \cdots = P_K = 25$ dBm
and the maximum sum power $P = 30$ dBm.
Under the assumption that the channel follows the circularly symmetric complex Gaussian distribution
with unit noise variance, i.e. $\bm{H}\sim \mathcal{C N} \left(\mathbf{0}, \bm{I}_{N_t K} \right)$,
the SNR is defined as ${\rm SNR} = 10 \log_{10} (\frac{P}{\sigma^2} )$, where noise power $\sigma^2$ varies with SNR.
All the results are obtained by averaging over 100 channel realizations.
Firstly, Fig. \ref{fig_sim_diag1} compares the capacity and MSE performance 
achieved by the CMD-based algorithm 
and the numerical CVX optimization\cite{refer_cvx} versus SNR.
It is clearly observed that the CMD-based algorithm
achieves almost the same capacity and MSE performance as the numerical CVX optimization, which demonstrates its global optimality.

Then, the amplitude-adjustable IRS-aided MIMO system is taken into account, where the BS equipped with $N_t=6$ antennas and $N_{rf} = 4$ radio-frequency (RF) chains communicates with the user equipped with $N_r=4$ antennas,  while an $8\times 8$ IRS is deployed to enhance the point-to-point communication.
The path loss setting is the same as that in \cite{Xing_KKT}
and other parameter settings are the same as those in the uplink MU-SIMO system.
Fig. \ref{fig_sim_diag2} illustrates that the capacity and MSE performance 
attained by the CMD-based algorithm 
and the numerical CVX optimization as the function of SNR.
We also find that the CMD-based algorithm and the numerical CVX optimization attain almost the same optimal performance.

\vspace{-3.5mm}
\subsection{Constant Modulus Constraints}
\vspace{-1mm}

In the point-to-point hybrid analog-digital MIMO system, the BS equipped with $N_t=6$ antennas and $N_{rf} = 4$ RF chains serves a single-antenna user equipped with $N_r=4$ antennas.
Other parameter settings are the same as those of the uplink MU-SIMO system.
We firstly demonstrate the effectiveness of the proposed novel AO algorithm in Fig. \ref{fig_sim_feasi_point},
where three sets of different feasible solutions satisfying constant modulus constraints are generated randomly to calculate the optimal solutions of capacity maximization and MSE minimization problems, respectively.
The same capacity and MSE performance are attained for different feasible solutions,
implying the stability and effectiveness of the proposed novel AO algorithm.
Next, Fig. \ref{fig_sim_hybrid} compares the capacity and MSE performance of the proposed novel AO algorithm and the hybrid beamforming algorithm in \cite{Wei_Yu_Hybrid} versus SNR.
Specifically, the authors of \cite{Wei_Yu_Hybrid} propose an element-wise BCD algorithm for optimizing the analog beamforming matrix.
Obviously, our proposed algorithm achieves almost the same performance as the benchmark schemes. 
Taking the capacity maximization problem as an example, Fig. \ref{fig_sim_conver} compares the convergence behavior of the proposed novel AO algorithm and the hybrid beamforming algorithm in \cite{Wei_Yu_Hybrid},
where SNR$=5$ dB.
As can be seen, the proposed novel AO algorithm shows a little higher capacity than the algorithm in \cite{Wei_Yu_Hybrid}.
Moreover, both the two algorithms converge within $10$ iterations, while the proposed algorithm has a faster speed than the algorithm in \cite{Wei_Yu_Hybrid}.

\begin{figure*}[!t]
  \vspace{-14mm}
	\centering
  \subfigure[The capacity and MSE performance comparison]
  {
  \includegraphics[width=0.31\textwidth]{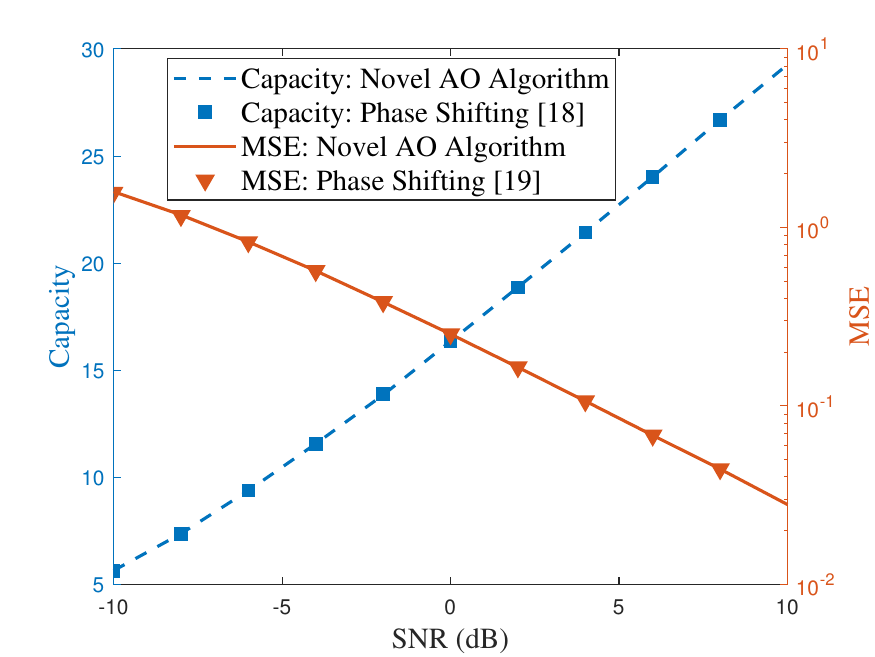} 
  \label{fig_sim_irs}
  }%
  \subfigure[Average CPU runtime comparison]
  {
  \includegraphics[width=0.31\textwidth]{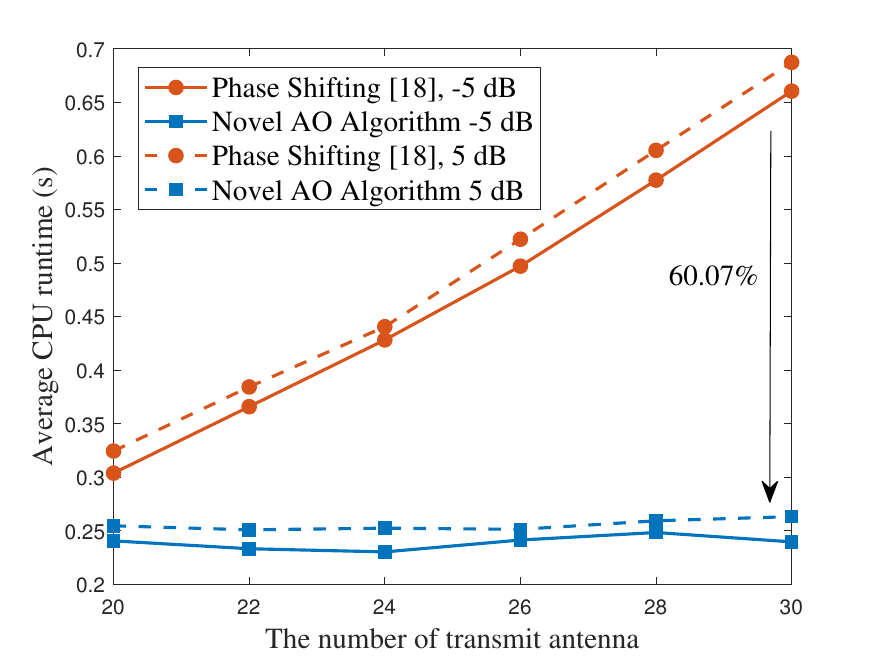}
	\label{fig_sim_runtime}
  }%
  \vspace{-2mm}
  \caption{The capacity and MSE performance comparison in the fully-passive IRS-aided MIMO system.}
  \label{fig_feasi_hybrid2}
  \vspace{-6mm}
\end{figure*}

Furthermore, we consider the fully-passive IRS-aided MIMO system with the same parameter settings as the amplitude-adjustable IRS-aided MIMO system.
In Fig. \ref{fig_sim_irs}, the capacity and MSE performance of the proposed novel AO algorithm are compared with that of the phase shifting algorithms in \cite{S_Zhang_IRS} and \cite{XinZhaoRIS}, respectively.
These two benchmark schemes adopt the element-wise BCD algorithm to optimize IRS reflection matrix for capacity maximization and MSE minimization problems, respectively.
It is obvious from Fig. \ref{fig_sim_irs} that the proposed novel AO algorithm is able to achieve the same performance as the benchmark schemes.

Finally, in order to demonstrate the low-complexity advantage of the proposed AO algorithm, 
Fig. \ref{fig_sim_runtime} compares its average CPU runtime with the phase shifting algorithm in \cite{S_Zhang_IRS} for the capacity maximization problem,
where $N_r = 8$. 
We firstly observe that the average CPU runtime of each  studied algorithm increases with  SNR, since more power resource need to be allocated for enhancing system performance. 
Furthermore, it is  seen that the average CPU runtime of the algorithm in \cite{S_Zhang_IRS} increases as the number of transmit antennas increases, 
since the dimension of the involved matrix inversion is the same as the number of antennas.
Whereas, the average CPU runtime of the proposed algorithm is almost the same, 
since it avoids the complicated matrix inversion.
Moreover, the proposed novel AO algorithm shows a sharp decrease of the average CPU runtime relative to the algorithm in \cite{S_Zhang_IRS}.
Clearly, there is $60.07\%$ CPU runtime decrement at $N_t=30$ and SNR$=-5$ dB,
implying the low-complexity advantage of this novel AO algorithm.

\vspace{-2mm}
\section{Conclusions}\label{sec_conclusion}
In this paper, we investigated complex matrix derivatives for two special matrices, i.e., 
diagonal structured matrices and  constant modulus structured matrices. 
Under the diagonal structure constraints, the optimal closed-form solutions of the  capacity maximization
problem, the  MSE minimization problem and their variants can be obtained using complex matrix derivatives. Whereas for constant modulus constraints, the
optimal solutions of these classical optimization problems are derived utilizing element-wise
phase derivatives. Further, in order to avoid the complicated matrix operations, we explore the
inherent structure of the element-wise phase derivatives, and develop a novel AO algorithm with
the aid of several arbitrary feasible solutions. Finally, numerical simulations demonstrate the
global optimality and low complexity of the proposed novel AO algorithm.

\begin{figure*}[ht] 
  \vspace{-12mm}
  \centering 
\begin{align}\label{general_g_hybrid}
  &\tan ( \angle g_{i,j}(\bm{\widehat X}_m) ) 
  = \frac{\Im \{g_{i,j}(\bm{\widehat X}_m) \} }{\Re \{g_{i,j}(\bm{\widehat X}_m) \} }\nonumber\\
 =&\frac{\Re \{A^{\rm{CL}}_{i,j} \} \Im \{\bm{\widehat X}_m \} +\Im \{A^{\rm{CL}}_{i,j} \}\Re \{ \bm{\widehat X}_m\} +\Im \{B^{\rm{CL}}_{i,j} \}\Re \{\bm{\widehat X}_m \} 
  -\Re \{B^{\rm{CL}}_{i,j} \} \Im \{\bm{\widehat X}_m \} + \Im \{C^{\rm{CL}}_{i,j} \}}
  {\Re \{A^{\rm{CL}}_{i,j} \} \Re \{\bm{\widehat X}_m \} -\Im \{A^{\rm{CL}}_{i,j} \} \Im \{ \bm{\widehat X}_m\} +\Re \{B^{\rm{CL}}_{i,j} \}\Re \{\bm{\widehat X}_m \} 
  +\Im \{B^{\rm{CL}}_{i,j} \} \Im \{\bm{\widehat X}_m \} + \Re \{C^{\rm{CL}}_{i,j} \}}\nonumber\\
  =& \frac{[\bm{w}_{i,j}]_1 \Im \{\bm{\widehat X}_m \} +[\bm{w}_{i,j}]_2\Re \{ \bm{\widehat X}_m\} +[\bm{w}_{i,j}]_4\Re \{\bm{\widehat X}_m \} 
  -[\bm{w}_{i,j}]_3 \Im \{\bm{\widehat X}_m \} + [\bm{w}_{i,j}]_5}
  {[\bm{w}_{i,j}]_1 \Re \{\bm{\widehat X}_m \} -[\bm{w}_{i,j}]_2 \Im \{ \bm{\widehat X}_m\} +[\bm{w}_{i,j}]_3\Re \{\bm{\widehat X}_m \} 
  +[\bm{w}_{i,j}]_4 \Im \{\bm{\widehat X}_m \} + 1}
  ,\ m=1,\cdots,5.
\end{align}
 \hrulefill 
\vspace{-5mm}
\end{figure*}

\vspace{-2.5mm}

\appendix

\section*{ }
\vspace{-6mm}
\subsection{Proof of Proposition 1}\label{proof_gen_form}
 Based on Sec. \ref{sec_const_app}, we can easily conclude that 
for both capacity maximization problems (i.e. \textbf{Prob.8}, \textbf{Prob.11}) with log-determinant functions and WMMSE minimization problems (i.e. \textbf{Prob.10}, \textbf{Prob.13}) with trace-linear and trace-quadratic functions,
the corresponding element-wise phase derivatives have the  same linear forms as in (\ref{linear_form}).
Whereas, for the MSE minimization problems (i.e. \textbf{Prob.9}, \textbf{Prob.12}) with trace-inverse functions,
the element-wise phase derivatives satisfy the conjugate linear forms in (\ref{conj_linear_form}). 

Furthermore, in order to determine the optimal solution from two zero-derivative points $\left\{\left[\bm{X}_{1}\right]_{i,j},\left[\bm{X}_{2}\right]_{i,j}\right\}$ satisfying $g_{i,j}(\bm{X})=0, \forall i,j$, we resort to the second-order  derivative of the corresponding objective function $f(\bm{X})$  w.r.t. $\left[\bm{\Theta}\right]_{i,j}$'s. Specifically, by taking the conjugate linear element-wise phase derivatives
 as an example, we  have 
\begin{align}
  \frac{\partial^2 f(\bm{X})}{\partial \left[\bm{\Theta}\right]^2_{i,j}}
  \! =\! \frac{\partial g_{i,j}(\bm{X})}{\partial \left[\bm{\Theta}\right]_{i,j}}
  \!=\! 2 j \Re\! \left\{\! A^{\rm{CL}}_{i,j}[\bm{X}]_{i,j}\!-\!B^{\rm{CL}}_{i,j}[\bm{X}]^*_{i,j} \!\right\}.
\end{align}
According to  the  optimization theory, it is readily inferred  that these two  zero-derivative points are  local minimum  when 
$\frac{\partial^2 f(\bm{X})}{\partial \left[\bm{\Theta}\right]^2_{i,j}}\geq 0$ holds. In contrast, they become 
local maximum when $\frac{\partial^2 f(\bm{X})}{\partial \left[\bm{\Theta}\right]^2_{i,j}}<0$ holds\cite{Boyd04}. This completes the proof.


\vspace{-3mm}
\subsection{Proof of Proposition 2}\label{proof_equ_method}
It follows from \textbf{Proposition 1} that the derivation of the optimal $\bm{X}_{i,j}$'s is based on $A^{\rm{CL}}_{i,j}$'s, $B^{\rm{CL}}_{i,j}$'s and $C^{\rm{CL}}_{i,j}$'s.
Thus, we firstly define a $5$-dimensional vector as follows.
\begin{align}\label{w_abc}
  &\bm{w}_{i,j} 
  =\!\left[ [\bm{w}_{i,j}]_1,[\bm{w}_{i,j}]_2,\cdots, [\bm{w}_{i,j}]_5 \right]^{\rm{T}}\\
  &=\!\! \left[\! \frac{\Re \{A^{\rm{CL}}_{i,j} \}}{\Re \{C^{\rm{CL}}_{i,j} \}}, \! \frac{\Im \{A^{\rm{CL}}_{i,j} \}}{\Re \{C^{\rm{CL}}_{i,j} \}}, \!
  \frac{\Re \{B^{\rm{CL}}_{i,j} \}}{\Re \{C^{\rm{CL}}_{i,j} \}},\!\frac{\Im \{B^{\rm{CL}}_{i,j} \}}{\Re \{C^{\rm{CL}}_{i,j} \}},\! \frac{\Im \{C^{\rm{CL}}_{i,j} \}}{\Re \{C^{\rm{CL}}_{i,j} \}} \!\right]^{\rm{T}}\!\!\!\!
,    \ \forall i,j.\nonumber
\end{align}
Moreover, by recalling \textbf{Prob.12}, $A^{\rm{CL}}_{i,j}$'s, $B^{\rm{CL}}_{i,j}$'s and $C^{\rm{CL}}_{i,j}$'s are closely related to the functions $g_{i,j}(\bm{X})$'s
associated with the original element-wise phase derivatives as (\ref{general_g_hybrid}), as shown at the top of the next page.
Note that the equations in (\ref{general_g_hybrid}) can be further rewritten as the following homogeneous linear equations, i.e.,
\begin{align}\label{general_g_hybrid2}
  &[\bm{w}_{i,j}]_1 \left(\Im \{\bm{\widehat X}_m \}-\tan ( \angle g_{i,j}(\bm{\widehat X}_m) ) \Re \{\bm{\widehat X}_m \} \right)\nonumber\\
+&[\bm{w}_{i,j}]_2 \left(\Re \{\bm{\widehat X}_m \}+\tan ( \angle g_{i,j}(\bm{\widehat X}_m) ) \Im \{\bm{\widehat X}_m \} \right)\nonumber\\
-&[\bm{w}_{i,j}]_3 \left(\Im \{\bm{\widehat X}_m \}+\tan ( \angle g_{i,j}(\bm{\widehat X}_m) ) \Re \{\bm{\widehat X}_m \} \right)\nonumber\\
+&[\bm{w}_{i,j}]_4 \left(\Re \{\bm{\widehat X}_m \}-\tan ( \angle g_{i,j}(\bm{\widehat X}_m) ) \Im \{\bm{\widehat X}_m \} \right)\nonumber\\
+&[\bm{w}_{i,j}]_5-\tan ( \angle g_{i,j}(\bm{\widehat X}_m) )=0,\ m=1,\cdots,5.
\end{align}
It follows from (\ref{general_g_hybrid2}) that the optimal $\bm{w}_{i,j}$'s can be obtained by jointly solving its involved five homogeneous linear equations, 
whose closed-form  structures are  further shown  in (\ref{general_equ}) and the proof of \textbf{Proposition 2} is completed.



\end{document}